\documentclass[onecolumn,aps,prb,showpacs,preprintnumbers,amsmath,amssymb]{revtex4}
\usepackage{epsfig}
\usepackage{graphicx}
\usepackage{dcolumn}
\usepackage{bm}

\renewcommand{\dag}{^{\dagger}}

\newcommand{\sx}{\sigma_x}
\newcommand{\sy}{\sigma_y}
\newcommand{\sz}{\sigma_z}

\newcommand{\De}{\Delta}

\newcommand{\dl}{\partial_{\ell}}

\newcommand{\LXR}{\langle\sigma_x\rangle}
\newcommand{\LZR}{\langle\sigma_z\rangle}

\newcommand{\bb}{\bar{b}}

\begin{document}
\title{Flow equations for Hamiltonians: Contrasting different approaches by using a numerically solvable model}

\author{T.~Stauber and A.~Mielke} \affiliation{Institut f\"ur Theoretische
  Physik, Ruprecht-Karls-Universit\"at Heidelberg, Philosophenweg 19,
  D-69120~Heidelberg, Germany } \date{\today}

\begin{abstract}
  To contrast different generators for flow equations for Hamiltonians 
  and to discuss the dependence of physical quantities on unitarily 
  equivalent, but effectively different initial Hamiltonians, a 
  numerically solvable model is considered which is structurally similar 
  to impurity models. By this we discuss the question of optimization for 
  the first time. A general truncation scheme is established that produces 
  good results for the Hamiltonian flow as well as for the operator flow.
  Nevertheless, it is also pointed out that a systematic and feasible
  scheme for the operator flow on the operator level is missing. For this, an
  explicit analysis of the operator flow is given for the first time. We 
  observe that truncation of the series of the observable flow after
  the linear or bilinear terms does not yield satisfactory results for
  the entire parameter regime as - especially close to resonances -
  even high orders of the exact series expansion carry considerable
  weight.
\end{abstract}
%

%

\maketitle 

\section{Introduction}

\subsection{Flow equations}

Eight years ago, G{\l}azek and Wilson \cite{Gla94} and independently 
Wegner \cite{Weg94} introduced a new non-perturbative method to
diagonalize, renormalize or simplify a given Hamiltonian.
Whereas in high energy physics the method is known as
``similarity transformations'', the term ``flow equations'' has been
established in the solid-state community.
The idea is conceptually simple: Instead of diagonalizing the
Hamiltonian of the system by a single unitary transformation, one
performs a continuous sequence of infinitesimal unitary
transformations and thus induces a flow on the system parameters. The
procedure is not constrained to specific symmetries nor to certain
parameter regimes - but is accessible to any system described by a
Hamiltonian. Thus, the method has been successfully applied to various
models of solid-state and nuclear physics.
Examples are dissipative quantum systems \cite{Keh96a,Keh96b,Keh97}, 
the electron-phonon problem
\cite{Len96,Rag99} or the Hubbard model \cite{Gro02}. 
For a recent review on the flow equation method see Ref. \onlinecite{Weg01}. 

The main advantage of the method is its flexibility. This is similar
to the numerical diagonalization of a given matrix: There are many
different possibilities to reach the goal. One is free to choose the basis
in which the diagonalization is performed and within a given basis
one is free to choose the concrete series of unitary transformations
that finally diagonalizes the matrix. Depending on the basis and
on the concrete series of unitary transformations, convergence may be
good or poor, numerical errors may be small or large.

Similarly, many different flow equations can be formulated
to diagonalize or simplify a given Hamiltonian.
Even though all different flow equations are equivalent and will
eventually lead to the same result, matters change as soon as
approximations are involved.  Typically one needs to cut the
hierarchy of newly generated interaction terms and then neglect
operators, which are assumed to be {\em irrelevant}. Yet, there is no
satisfactory definition for {\em irrelevant operators} within the flow
equation approach. Whether or not a contribution is irrelevant depends
on the initial Hamiltonian and on the goal on wants to reach.

Usually, approximations were justified when certain sum rules, mostly
stemming from the invariance of commutation relations during the
unitary flow, hold exactly or at least asymptotically \cite{Keh97,Rag99}. In
addition, exact relations between static and dynamic properties - as
the generalized Shiba relation in the case of the spin-boson model \cite{Sas90} - can serve as justification for prior approximations \cite{Keh97}.  A
general consistency check lies in the explicit investigation of the
flow of the neglected operators.

So far, a detailed discussion on optimization of flow equations is missing. With this work, we want to start to fill this gap by addressing the following questions: E.g., any initial Hamiltonian $H$ implicitly depends on a number of parameters $H=H(\psi,\theta,...)$ where the parameters are associated to certain unitary transformations $U_1(\psi)$, $U_2(\theta)$, $...$ . Can the parameters $\psi$, $\theta$, $...$ be chosen such that a given flow equation scheme yields optimal results? A second variation that will be discussed lies in the arbitrary definition of the ``diagonal'' Hamiltonian, $H_0$ - as mentioned above. Will different $H_0$ yield similar results for physical quantities and is there an optimal $H_0$ for all physical quantities - or does the optimal $H_0$ depend on the physical quantity under scrutiny? 

Another fundamental question associated with the flow equation approach is connected to the observable flow and has not been discussed in depth yet, either. For this, we note that in order to take advantage of the simple structure of the fixed point Hamiltonian, the observable has to be transformed as well - by the same sequence of unitary transformations that diagonalized the Hamiltonian. Since usually the continuous transformation is designed such that the diagonalization of the Hamiltonian is ``optimal'', the observable flow is more likely to suffer from uncontrolled approximations. We will address the question if there is a scheme that optimizes both, Hamiltonian and observable flow and also compare the observable flow on the operator level. 

To do so, we will not proceed systematically but we will address these questions more specifically. Namely, we will consider an explicit model which is structurally similar to
dissipative impurity models - but still exactly solvable via numerical
diagonalization. We will call this model the Rabi model and it is presented in the next subsection. We use this model to test different approximation schemes
and compare the results with the numerically exact solution. This
strategy was first pursued by Richter in his diploma work
\cite{Ric97}. We extend his work in various directions. One point is
to investigate Hamiltonians where the
reflection symmetry is broken. This is important if one wants to
understand the mechanism of phase transition as being observed in the
spin-boson model \cite{Leg87}.

In Sec. II, we develop a general truncation scheme
which yields good results over a wide range of the parameter space.
Furthermore, we present a particular truncation scheme which leaves
the Hamiltonian form-invariant during the flow.  The question of the
invariance of the flow equations with respect to the particular choice
of initial Hamiltonians, provided that they only vary by a unitary
transformation, is discussed.  As a criterion for the quality of the
flow equations, we look at the ground-state energy as function of the
bias as an example for the flow of a parameter of the Hamiltonian.
 In Sec. III, we give a thorough discussion about the flow of observables.
As reference we will not only investigate the expectation value of
observables, but also compare the flow equation result with the exact
solution {\em on the operator level} for the first time. For this, an expansion of the operator
into a basis of normal ordered bosonic operators is given.
 In Sec. IV, we close with general remarks and conclusions.

\subsection{The Rabi Hamiltonian}

The specific model we use for our discussion of various realizations
of flow equations and various approximation schemes is the 
spin-boson Hamiltonian with only one mode, which we will call the
Rabi model in order to distinguish it from the spin-boson model with
an arbitrary number of modes. The Hamiltonian is given by
\begin{align}
  \label{RabiHamiltonian} 
  H=-\frac{\De_0}{2}\sigma_x + \frac{\epsilon_0}{2}\sigma_z+ \omega_0
  b\dag b+\sigma_z\frac{\lambda_0}{2}(b+b\dag) +E_0\quad.
\end{align}
Here, $b^{(\dagger)}$ denotes the bosonic  degree of freedom
and $\sigma_i$ with $i=x,y,z$ are the Pauli spin matrices. They
obey the canonical commutation relation $[b,b\dag]=1$ and the
spin-$1/2$ algebra $[\sigma_i,\sigma_j]=2i\epsilon_{ijk}\sigma_k$. 
Since there is only one mode present, a numerical diagonalization is
feasible by truncating the bosonic Hilbert space after $n$ bosonic
excitations with some fixed value of $n$. 

The model was first introduced in the context of spontaneous emission
and absorption of atoms and due to its long history 
there exists an enormous amount
of work that has already been published on this model. 
It is impossible to review or cite all these papers - a good
overview may be found in the paper by Graham {\it et al} \cite{GH84}. 
The model has also been discussed in connection with quantum chaos
\cite{Cibils91, Cibils92} and extensions of it can serve 
for the description of optical phonons interacting 
with two-level systems or quantum dots within 
a solid-state matrix \cite{Sta00}. In the context of flow equations,
the model has been discussed by Mielke \cite{Mielke98} using a
set of flow equations that preserve the banded structure of
the Hamiltonian. In the present work we focus on low energy properties
of the Rabi model. Ref. \onlinecite{Mielke98} is in some sense complementary
to the present work since there the high energy modes were discussed.

In the present work we are interested
in general properties of the flow equation method. The reasons for us
to investigate the Rabi model lie in the fact that it couples a two-level
system to a ``bath'' resembled by the bosonic degree of freedom. And
since we only consider one mode, the system is still exactly solvable
via numerical diagonalization.

%
%

\section{Flow of the Hamiltonian}

\subsection{Setting up the basis}

In order to diagonalize the Hamiltonian (\ref{RabiHamiltonian}), 
we will perform a continuous unitary transformation. The flow
equations are generated by the anti-hermitian operator $\eta$ which is
canonically given by $\eta=[H_0,H]$, where $H_0$ defines the diagonal
Hamiltonian \cite{Weg94}. Different choices of $\eta$ are possible 
as well. The flow equations are of the form
\begin{align}
\frac{dH}{d\ell}=[\eta,H]\quad,
\end{align}
where both $H$ and $\eta$ depend on the flow parameter $\ell$.
The choice $\eta=[H_0,H]$ is likely to decouple the fermionic system
from the bosonic system, and the fixed point Hamiltonian $H(\ell=\infty)$
is then basically given by $H_0^*$ where the asterisk indicates that
the parameters of the initial diagonal Hamiltonian are in general
renormalized. For a brief introduction, we refer
to Appendix \ref{IndependentBoson}, which treats the Rabi model with
$\Delta=0$ and motivates the approach given here.

Obviously, different choices for $H_0$ can lead to different flow
equations.  Another ambiguity stems from the fact that the initial
Hamiltonian may differ by a unitary transformation. If we restrict
ourselves to orthogonal transformations in the two-dimensional Hilbert
space and to simple translations in the bosonic Hilbert space, i.e.
\begin{align}
  U_S=\begin{pmatrix}
    \cos\frac{\psi}{2}&\sin\frac{\psi}{2}\\
    -\sin\frac{\psi}{2}&\cos\frac{\psi}{2}
\end{pmatrix}\quad,\quad U_B=
\exp(\theta\frac{\lambda_0}{2\omega_0}(b-b\dag))\quad,
\end{align}
the general initial Hamiltonian with respect to the Hamiltonian (\ref{RabiHamiltonian}) is given by
\begin{align}
  \label{InitialHamiltonianUnitary}
  H&=-\frac{\De'}{2}\sigma_x + \frac{\epsilon'}{2}\sigma_z +\omega_0
  b\dag b
  +\frac{\lambda^e}{2}(b+b\dag)+\sigma_x\frac{\lambda^x}{2}(b+b\dag)
  +\sigma_z\frac{\lambda^z}{2}(b+b\dag)+E'\quad,
\end{align}
where we introduced the following parameters:
\begin{eqnarray}
&\De'=\De_0\cos\psi+\tilde{\epsilon}\sin\psi\quad,\quad
\epsilon'=\tilde{\epsilon}\cos\psi-\De_0\sin\psi&\\
&\lambda^e=\theta\lambda_0\quad,\quad\lambda^x=-\sin\psi\lambda_0\quad,\quad
\lambda^z=\cos\psi\lambda_0&\\\label{InitialEnergyShift}
&E_0'=E_0+\theta^2\frac{\lambda_0^2}{4\omega_0}\quad,\quad\tilde{\epsilon}
=\epsilon_0+\theta\frac{\lambda_0^2}{\omega_0}&
\end{eqnarray}

As was mentioned above, different generators and different unitarily
equivalent Hamiltonians will lead to the same physical results if no
approximations are involved. But the above model is not solvable
analytically and therefore approximations become necessary. In the
special case of the Rabi model the flow equations will generate an
infinite series of new coupling terms which cannot be summed
up formally to yield a closed expression \cite{close}.

In this section we will first only take coupling terms into account
which are linear in the bosonic operators and have real coefficients.
This means that with respect to the initial Hamiltonian 
(\ref{InitialHamiltonianUnitary}), only the term $i\sy(b-b\dag)$ will be
newly generated which resembles the lowest order of the polaron transformation (see e.g. Ref. \onlinecite{Leg87}). 
Using a generator which is not of the simple form
$\eta=[H_0,H]$, we will also discuss flow equations which leave the
initial Hamiltonian form-invariant. We are able to show
analytically that the fixed point of the flow equation is independent
with respect to the (distinguished) unitary transformation.

\subsection{Flow equations with respect to the Canonical Generator} 
\label{Unshifted_Rabi}
In the following subsection we will discuss flow equations which are
obtained by employing the canonical generator $\eta=[H_0,H]$. This
gives rise to new interaction terms. The truncated Hamiltonian shall
be given by
\begin{align}
  \label{TrunkHamiltonianUnShifted}
  H&=-\frac{\De}{2}\sigma_x + \frac{\epsilon}{2}\sigma_z +\omega_0
  b\dag b
  +\frac{\lambda^e}{2}(b+b\dag)+\sigma_x\frac{\lambda^x}{2}(b+b\dag)
  +i\sigma_y\frac{\lambda^y}{2}(b-b\dag)
  +\sigma_z\frac{\lambda^z}{2}(b+b\dag)+E\quad,
\end{align}
where all parameters but the bath energy $\omega_0$ are explicitly $\ell$-dependent.  The above
Hamiltonian represents the most general Hermitian operator which
includes all possible interaction terms acting on the underlying
Hilbert space up to linear bosonic operators with real coefficients.

The flow shall be governed by the generator
\begin{align}
\label{GeneratorUnShifted}
\eta&=i\sy \eta^{0,y}+\eta^e(b-b\dag)+\sx \eta^x(b-b\dag)+i\sy
\eta^y(b+b\dag)+\sz \eta^z(b-b\dag)\\\notag &\equiv
\hat{\eta}^{0,y}+\hat{\eta}^e+\hat{\eta}^x+\hat{\eta}^y+\hat{\eta}^z\quad,
\end{align}
where the parameters $\eta^{0,y}$, $\eta^e$, $\eta^x$, $\eta^y$, and
$\eta^z$ are $\ell$-dependent and will be specified later.

The above generator represents the most general anti-Hermitian
operator which includes all possible operators acting on the
underlying Hilbert space up to linear bosonic operators with real
coefficients.
\subsubsection{Setting up the flow equations}
The commutator $[\eta,H]$ yields the following contributions:
\begin{align}
  [\hat{\eta}^{0,y},H]&=-\sz\De \eta^{0,y}-\sx\epsilon \eta^{0,y}+\sz
  \eta^{0,y}\lambda^x(b+b\dag)
  -\sx \eta^{0,y}\lambda^z(b+b\dag)\\
  [\hat{\eta}^e,H]&=\eta^e\omega_0(b+b\dag)+\eta^e\lambda^e
  +\sx\eta^e\lambda^x+\sz\eta^e\lambda^z\\
  [\hat{\eta}^x,H]&=-i\sy\epsilon \eta^x(b-b\dag)
  +\sx \eta^x\omega_0(b+b\dag)\notag\\
  &+\sx\eta^x\lambda^e+\eta^x\lambda^x-\sz \eta^x\lambda^y(b-b\dag)^2
  -i\sy \eta^x\frac{\lambda^z}{2}\{(b-b\dag),(b+b\dag)\}\label{Oetax}\\
  [\hat{\eta}^y,H]&=-\sz\De \eta^y(b+b\dag)-\sx\epsilon
  \eta^y(b+b\dag)
  +i\sy \eta^y\omega_0(b-b\dag)\notag\\
  &+\sz \eta^y\lambda^x(b+b\dag)^2
  +\eta^y\lambda^y-\sx \eta^y\lambda^z(b+b\dag)^2\\
  [\hat{\eta}^z,H]&=-i\sy\De \eta^z(b-b\dag)+\sz
  \eta^z\omega_0(b+b\dag)
  +\sz\eta^z\lambda^e\notag\\
  &+i\sy \eta^z\frac{\lambda^x}{2}\{(b-b\dag),(b+b\dag)\}+\sx
  \eta^z\lambda^y(b-b\dag)^2 +\eta^z\lambda^z\label{Oetaz}
\end{align}

$\{.,.\}$ denotes the anti-commutator. As can be seen in 
(\ref{Oetax}) - (\ref{Oetaz}), the flow equations generate terms which
are bilinear in the bosonic operators and we will need to find a
suitable procedure how to include these terms in the flow. Kehrein,
Mielke, and Neu \cite{Keh96a} proposed to neglect these terms after
normal ordering them with respect to a bilinear bosonic Hamiltonian.
Since we allow the initial Hamiltonian to differ by a shift in the
bosonic operators, we need to include this generalization also in the
normal ordering procedure, i.e. we will normal order with respect to
the shifted bosonic mode
\begin{align}
  \bb\equiv b+\frac{\delta}{2}\quad,
\end{align}
with the linear shift $\delta$ to be determined later.  To close the
flow equations we will thus neglect the normal ordered operators
\begin{align}
\label{Neglect_Rabi}
\mathcal{O}_1&=-\sigma_x \eta^y\lambda^z:(\bb +\bb \dag)^2:\quad,\quad
\mathcal{O}_2=\sigma_z \eta^y\lambda^x:(\bb +\bb \dag)^2:\quad,\\
\mathcal{O}_3&=\sigma_x \eta^z\lambda^y:(\bb -\bb \dag)^2:\quad,\quad
\mathcal{O}_4=-\sigma_z \eta^x\lambda^y:(\bb -\bb \dag)^2:\quad,\\
\mathcal{O}_5&=i\sigma_y
(\eta^z\frac{\lambda^x}{2}-\eta^x\frac{\lambda^z}{2}) :\{(\bb -\bb
\dag),(\bb +\bb \dag)\}:\quad.
\end{align}
Normal ordering is now defined as $:(\bb +\bb \dag)^2:\equiv(\bb +\bb
\dag)^2-1_n$, with $1_n\equiv\langle(\bb +\bb \dag)^2\rangle=1+2n$,
and $n=(e^{\beta\omega_0}-1)^{-1}$ being the Bose factor. Notice that
the temperature enters in the Hamiltonian flow through normal
ordering. In the following we will only consider $T=0$, i.e. $1_n=1$,
but we will nevertheless keep track of this distinction.

Like in the case of flow equations for impurity systems \cite{Keh97},
the above truncation scheme has the effect that the bosonic energy
$\omega_0$ is not being renormalized during the flow. 

With $\frac{dH}{d\ell}=[\eta,H]$ we obtain the following flow equations:
\begin{align}
\label{Sym_Rabi_De}
\dl \De&= 2\epsilon \eta^{0,y}-2\eta^e\lambda^x-2\eta^x\lambda^e
+2(\eta^z\lambda^y+\eta^y\lambda^z)1_n-2\eta^y\lambda^z\delta^2\\
\dl \epsilon&=-2\De \eta^{0,y}+2\eta^z\lambda^e+2(\eta^y\lambda^x
+\eta^x\lambda^y)1_n+2\eta^e\lambda^z-2\eta^y\lambda^x\delta^2
\quad,\quad\dl  \lambda^e=2\eta^e\omega_0\\
\dl \lambda^x&=-2\epsilon
\eta^y+2\eta^x\omega_0-2\eta^{0,y}\lambda^z+4\eta^y\lambda^z\delta\quad,
\quad \dl \lambda^y=-2\De \eta^z-2\epsilon \eta^x+2\eta^y\omega_0
-2\eta^z\lambda^x\delta+2\eta^x\lambda^z\delta\\
\label{Sym_Rabi_E}
\dl \lambda^z&=-2\De
\eta^y+2\eta^z\omega_0+2\eta^{0,y}\lambda^x-4\eta^y\lambda^x\delta\quad,\quad
\dl E=\eta^e\lambda^e+\eta^x\lambda^x+\eta^y\lambda^y+\eta^z \lambda^z
\end{align} 
With $\lambda^e=\eta^e=0$, an obvious invariant is given by
$\text{Inv}=\De^2+\epsilon^2+{\lambda^x}^2
+{\lambda^y}^2+{\lambda^z}^2-4E\omega_0$.  To investigate the flow
equations further, one has to specify the constants and initial
conditions. To do so we will choose different diagonal Hamiltonians
$H_0$, and we will contrast the resulting flow equations by means of
the ground-state energy of the system.

\subsubsection{Determining the Canonical Generator}

An obvious choice for the diagonal Hamiltonian is given by
$H_0=-\frac{\De}{2}\sigma_x + \omega_0 b\dag b$.  The canonical
generator $\eta=[H_0,H]$ is of the form
(\ref{GeneratorUnShifted}) with $\eta^{0,y}=\De\epsilon/2$,
$\eta^e=-\omega_0\lambda^e/2$, $\eta^x=-\omega_0\lambda^x/2$,
$\eta^y=(\De\lambda^z-\omega_0\lambda^y)/2$ and
$\eta^z=(-\omega_0\lambda^z+\De\lambda^y)/2$.  We will refer to the
flow equations with this particular choice of $\eta$ as Version
a.

Another choice for the diagonal Hamiltonian is given by
$H_0=\frac{\epsilon}{2}\sigma_z + \omega_0 b\dag b$.  The canonical
generator $\eta=[H_0,H]$ is of the form
(\ref{GeneratorUnShifted}) with $\eta^{0,y}=-\De\epsilon/2$,
$\eta^e=-\omega_0\lambda^e/2$,
$\eta^x=(-\omega_0\lambda^x+\epsilon\lambda^y)/2$,
$\eta^y=(\epsilon\lambda^x-\omega_0\lambda^y)/2$ and
$\eta^z=-\omega_0\lambda^z/2$.  We will refer to the flow equations
with this particular choice of $\eta$ as Version b.

The third choice for the generator which we will investigate in the
following combines the two previous choices, i.e $\eta=[H_0,H]$ with
$H_0=-\frac{\De}{2}\sigma_x + \frac{\epsilon}{2}\sigma_z +\omega_0
b\dag b$. The canonical generator $\eta$ is of the form 
(\ref{GeneratorUnShifted}) with
$\eta^{0,y}=0$, $\eta^e=-\omega_0\lambda^e/2$,
$\eta^x=(-\omega_0\lambda^x+\epsilon\lambda^y)/2$,
$\eta^y=(\De\lambda^z+\epsilon\lambda^x-\omega_0\lambda^y)/2$ and
$\eta^z=(-\omega_0\lambda^z+\De\lambda^y)/2$.  We will refer to the
flow equations with this particular choice of $\eta$ as Version
c.

There are other possibilities for the diagonal Hamiltonian which
include coupling terms. We could e.g. choose $H_0=\omega_0b\dag
b+\sigma_z\frac{\lambda^z}{2}(b+b\dag)$, since this Hamiltonian is
also exactly solvable, see Appendix \ref{IndependentBoson}. Another
possibility is to choose the Jaynes-Cummings Hamiltonian as $H_0$
\cite{Jay63} (see also Appendix \ref{RabiPertubation}), which was done
by Richter \cite{Ric97}. In this work though, we want to confine
ourselves to the versions given above.

\subsubsection{Determining the Bosonic Shift}

We now want to determine the newly introduced bosonic shift $\delta$.
The procedure is not unambiguous, but we are led by formally
diagonalizing the Hamiltonian as follows:
\begin{align}
  H&=-\frac{\De'}{2}\sigma_x +\frac{\epsilon'}{2}\sigma_z
  +\omega_0(b\dag +\sum_j\sigma_j\frac{\lambda^j} {2\omega_0}
  +i\sigma_y\frac{\lambda^y}{2\omega_0})
  (b+\sum_j\sigma_j\frac{\lambda^j} {2\omega_0}
  -i\sigma_y\frac{\lambda^y}{2\omega_0})+E'\quad,
\end{align}
with $\De'\equiv\De+\frac{\lambda^e\lambda^x}{\omega}
-\frac{\lambda^y\lambda^z}{\omega}$,
$\epsilon'\equiv\epsilon-\frac{\lambda^e\lambda^z}{\omega}
-\frac{\lambda^x\lambda^y}{\omega}$, and $E'\equiv E
-\frac{\lambda^i\lambda^i}{4\omega}
-\frac{\lambda^y\lambda^y}{4\omega}$ and summation is over $j=e,x,z$
with $\sigma_e\equiv1$. Decoupling the fermionic and bosonic Hilbert
space, we thus obtain the $\ell$-dependent shift
\begin{align} 
\label{BosonicShift}
\delta=\sum_j\langle\sigma_j\rangle\frac{\lambda^j}{\omega_0}\quad.
\end{align}
The fermionic expectation values can be evaluated directly with
respect to the effective Hamiltonian
$H^p=-\frac{\De'}{2}\sigma_x+\frac{\epsilon'}{2}\sigma_z$ to yield
\begin{align}
\label{EVdirectly}
\langle\sigma_x\rangle={\De'}/{R'}\quad,\quad
\langle\sigma_z\rangle=-{\epsilon'}/{R'}\quad,\quad \text{with
  }{R'}^2\equiv{\De'}^2+{\epsilon'}^2.
\end{align}
\\
There is also a self-consistent possibility to determine the system
expectation values. For that we will formulated the Hamiltonian with
respect to the shifted mode $\bb=b+\delta/2$.  The renormalized
``one-particle'' parameters are then given by
\begin{align}
\label{EVselfcon}
\bar{\De}\equiv&\De+\lambda^x\delta\quad,\quad
\bar{\epsilon}\equiv\epsilon-\lambda^z\delta\quad.
\end{align}
Evaluating the system parameters now with respect to the system
Hamiltonian $H^p=-\frac{\bar{\De}}{2}\sigma_x +
\frac{\bar{\epsilon}}{2} \sigma_z$ and still assuming the bosonic
shift as given in Eq. (\ref{BosonicShift}), we obtain the following
self-consistent equations:
\begin{align}
\label{SelfconsistingSet}
\langle\sigma_x\rangle=\bar{\De}/\bar{R}\quad,\quad
\langle\sigma_z\rangle=-\bar{\epsilon}/\bar{R}\quad,\quad \text{with
  }\bar{R}^2\equiv\bar{\De}^2+\bar{\epsilon}^2\quad.
\end{align} 

In this work, we will restrict our investigation to the bosonic shift of (\ref{BosonicShift}) and to these two procedures of determining the fermionic expectation values.
But there are other possibilities of evaluating the bosonic shift or the
expectation values. One way is e.g. to couple the flow of the system
parameters with the flow of the observable by imposing that a certain
sum rule holds {\em exactly} (see next section). 
This condition will determine the
bosonic shift.  In the next section we show that the sum rule for the
$x$- and $z$-component of the Pauli matrices is quadratic in the
bosonic shift. But since we restricted ourselves to real shifts, there
might be no solution. Even if we allowed imaginary coefficients in the
evolution of the Hamiltonian, a solution would not be guaranteed since the
sum rule would then relate the complex shift $\delta$ with its complex
conjugate $\delta^*$. Numerical investigations indicated that the
bosonic shift $\delta$ cannot be chosen such that a certain sum rule
holds exactly.  It is left open, how this effects the stability and
reliability of the flow equation approach.

Finally, we want to point out that the procedure of determining the expectation values can significantly alter the behavior of the flow equations. In case of the spin-boson model it is shown \cite{StaDiss}, that an infinitesimal bias resembles a relevant perturbation, i.e. $\dl \epsilon\propto\epsilon$ for small $\ell$, if one chooses the expectation values directly whereas it resembles a irrelevant perturbation ($\dl \epsilon\propto-\epsilon$) if one chooses the self-consistent scheme.
  
\subsubsection{Numerical Results}

\begin{figure}[t]
  \begin{center}
    \epsfig{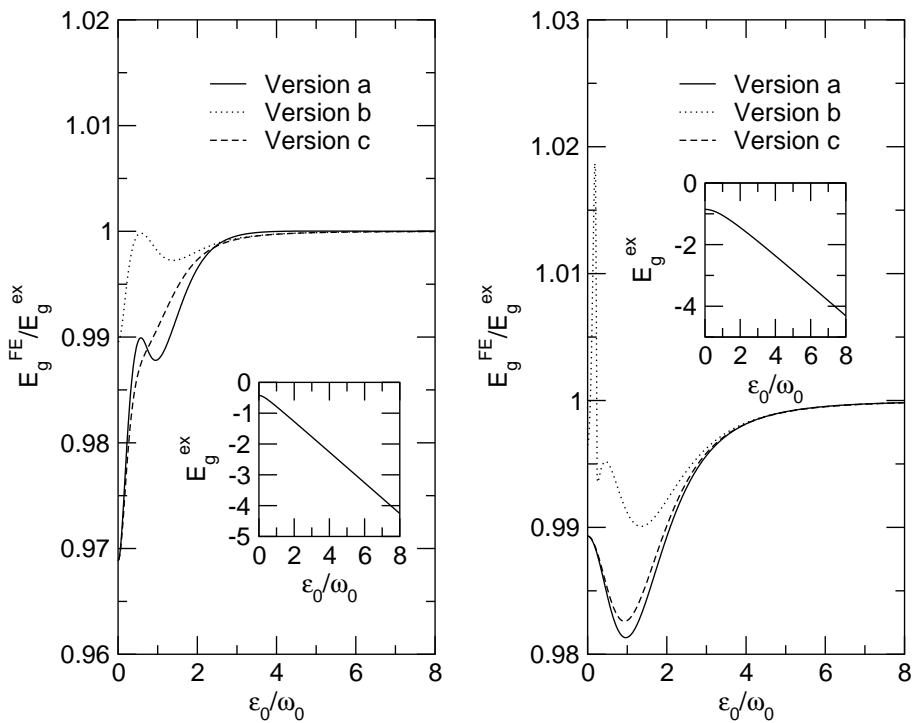}
    \caption{The ground-state energy $E_g^{FE}$ obtained by different 
      canonical generators with $\psi=0$ and $\theta=0$ for
      $\De_0/\omega_0=0.5$ (left hand side) and $\De_0/\omega_0=1.5$
      with $\lambda_0/\omega_0=1$ as a function of the bias
      $\epsilon_0$ relative to the exact ground-state energy
      $E_g^{ex}$, shown in the panel. The bosonic shift is set zero
      throughout the flow, i.e. $\delta=0$ for all $\ell$.}
    \label{Abb1}
\end{center}
\end{figure}
\begin{figure}[t]
  \begin{center}
    \epsfig{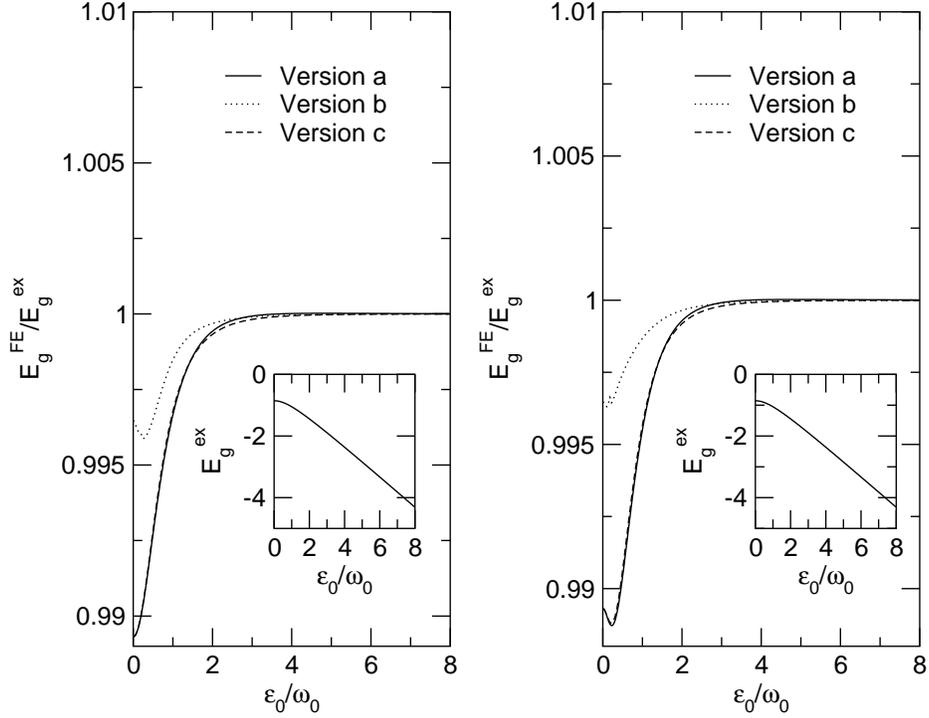}
    \caption{The ground-state energy $E_g^{FE}$ obtained by different 
      canonical generators with $\psi=0$ and $\theta=0$ for
      $\De_0/\omega_0=1.5$ with $\lambda_0/\omega_0=1$ as function of
      the bias $\epsilon_0$ relative to the exact ground-state energy
      $E_g^{ex}$, shown in the panel. The expectation values for the
      bosonic shift
      $\delta=\sum_j\langle\sigma_j\rangle\lambda^j/\omega_0$ were
      evaluated directly according to Eqs. (\ref{EVdirectly}) (left
      hand side) and self consistently according to Eqs.
      (\ref{SelfconsistingSet}) (right hand side).}
\label{Abb2}
\end{center}
\end{figure}
\begin{figure}[t]
  \begin{center}
    \epsfig{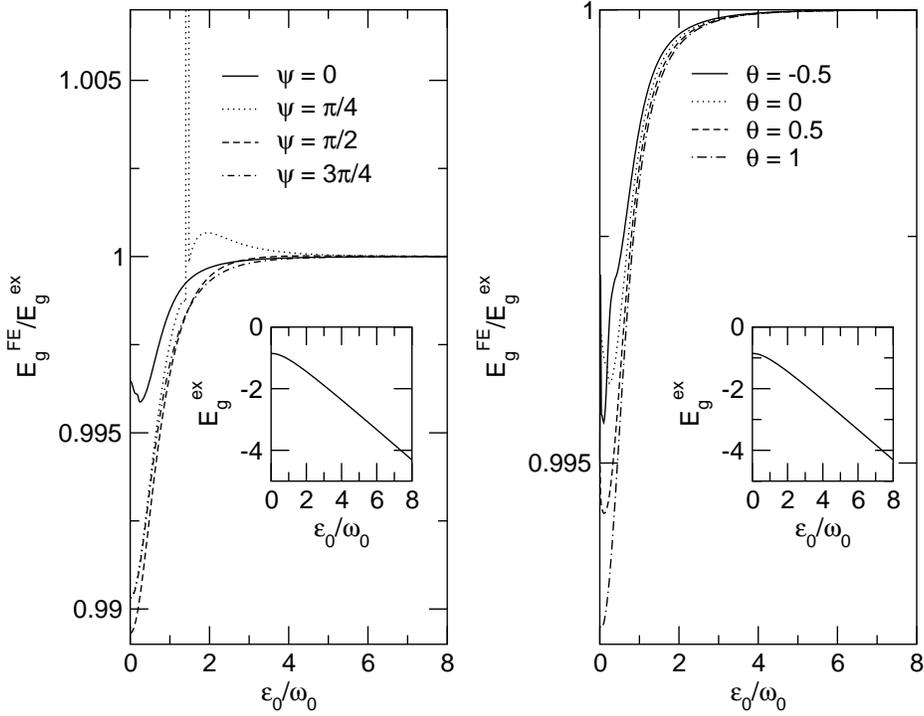}
    \caption{The ground-state energy $E_g^{FE}$ obtained by the canonical 
      generator of Version b for $\De_0/\omega_0=1.5$ and
      $\lambda_0/\omega_0=1$ as a function of the bias $\epsilon_0$
      relative to the exact ground-state energy $E_g^{ex}$, shown in
      the panel. The expectation values for the bosonic shift
      $\delta=\sum_j\langle\sigma_j\rangle\lambda^j/\omega_0$ were
      evaluated directly according to Eqs. (\ref{EVdirectly}). The
      initial values were $\theta=0$ and various $\psi$ (left hand
      side) and $\psi=0$ and various $\theta$ (right hand side).}
    \label{Abb3}
\end{center}
\end{figure}

We want to analyze the quality of the above flow equations by means of
the ground-state energy $E_g$ of the system as a function of the
external bias $\epsilon_0$. These results are compared with the
numerically exact solution obtained via numerical diagonalization.
Since the bosonic mode is left un-renormalized, the energy scale is
given by $\omega_0$. For the coupling constant we choose
$\lambda_0=\omega_0$, i.e. we are not in the perturbative regime.

We will first consider the flow of the initial Hamiltonian with
$\theta=0$ and $\psi=0$. We will also set $\delta=0$ for all $\ell$.
In Fig. \ref{Abb1} the ground-state energies $E_g^{FE}$ obtained from
the different canonical generators are shown.
Calculations are done for two different tunnel-matrix elements
$\Delta_0/\omega_0=0.5$ (left hand side) and $\Delta_0/\omega_0=1.5$
(right hand side), the first below and the second above resonance.
Resonance in the un-perturbed system is defined by $\De_0/\omega_0=1$.
All results are in good agreement with the numerically exact solution.
Still, differences occur in the non-trivial regime where the bias
$\epsilon_0$ is below or around the energy scale given by $\omega_0$.
In the panels, the exact ground-state energies $E_g^{ex}$ are
displayed.

We now turn to the flow equations obtained by employing the
generalized normal ordering procedure, i.e. we set
$\delta=\sum_j\langle\sigma_j\rangle\lambda^j/\omega_0$. The results
for the different generators are shown in Fig. \ref{Abb2}. The
expectation values are determined directly according to Eqs.
(\ref{EVdirectly}) (left hand side) and self-consistently according to
Eqs. (\ref{SelfconsistingSet}) (right hand side). There is a systematic
improvement to the results of Fig. \ref{Abb1}, where $\delta$ was set zero for
all $\ell$. The best results are obtained by the generator of
Version b and determining the expectation values self-consistently.

Finally, we want to investigate the dependence of the flow equations
on the unitarily equivalent, but different representations of the
initial Hamiltonian, labeled by $\psi$ and $\theta$. For this we choose the
generator of Version b and the bosonic shift of (\ref{BosonicShift}) with the direct evaluation of the expectation
values according to Eqs. (\ref{EVdirectly}). On the left hand side of
Fig. \ref{Abb3} we vary $\psi$ with $\theta=0$; on the right hand side
of Fig. \ref{Abb3} we vary $\theta$ with $\psi=0$.

As can be seen, there are differences with respect to the initial
Hamiltonian. For $\psi=\pi/4$, there is a big deviation from the exact
value in a small region around $\epsilon_0\approx1.5$ with a maximum
of $1.2$. In this region the fixed point Hamiltonian $H(\ell=\infty)$ varies
from the ``normal'' fixed point Hamiltonian and the ground-state energy
is mostly determined by $E(\ell=\infty)$. This is also the case for 
$\theta\leq-1$ (not shown) where the regions of large deviations 
depend on $\theta$.
Still, we observe a certain invariance with respect to the initial
Hamiltonian keeping the crude truncation scheme in mind.

From the considered parameters, the best results are obtained for
$\psi=0$ and $\theta=-0.5$. Of course, it would be desirable to give
an objective scheme how to choose the representation of the initial
Hamiltonian that yields the best result for the ground-state energy.
This had to be left open.

\subsection{Flow equations with respect to a Form-Invariant Flow}

As was mentioned above, the canonical generator $\eta=[H_0,H]$, in general, gives rise to new interaction terms. In order to avoid this complication, Kehrein,
Mielke, and Neu pursued a different strategy to set up the flow equations, namely they chose the generator $\eta$ such that the Hamiltonian remains form-invariant. 
To assure that the initial Hamiltonian of Eq. (\ref{RabiHamiltonian})
remains form-invariant, we set $\delta=0$, and the constants of the
generator of Eq. \ref{GeneratorUnShifted} have to satisfy the
following relations:
\begin{eqnarray}
\label{syp_Rabi}
\eta^e=0\quad,\quad-\epsilon \eta^y+\eta^x\omega_0-\eta^{0,y}\lambda^z
=0\quad,\quad
-\De \eta^z-\epsilon \eta^x+\eta^y\omega_0=0
\end{eqnarray}
This guarantees that $\lambda^e$, $\lambda^x$ and $\lambda^y$ are not
being generated.  With these relations, the parameters are defined up
to a common factor $f$.  If one chooses $\eta^z=-\omega_0\lambda^zf/2$, one
finds $\eta^{0,y}=\epsilon\De f/2$, $\eta^e=0$, $\eta^x=0$ and
$\eta^y=-\De\lambda^zf/2$.  With this choice, all neglected operators
except of $\mathcal{O}_1$ vanish. One obtains the following coupled
differential equations:
\begin{align}   
\label{Version1d}   
\begin{split}   
  \dl \De&=-\De{\lambda^z}^2f1_n+\De\epsilon^2f\quad,\quad
  \dl  \epsilon=-\epsilon\De^2f\\
  \dl \lambda^z&=\lambda^z(\De^2-\omega_0^2)f\quad,\quad \dl
  E=-\omega_0{\lambda^z}^2f/2
\end{split}
\end{align}
For the numerical calculations, we set $f=1$ and refer to this set of
flow equations as Version d.


We want to consider the form-invariant flow after having performed a
unitary transformation on the two-dimensional Hilbert space which
diagonalizes $H^p=-\frac{\De_0}{2}\sigma_x +
\frac{\epsilon_0}{2}\sigma_z\to\frac{R}{2}\sigma_z$ with
$R^2=\De_0^2+\epsilon_0^2$. This is achieved by choosing
$\tan\psi=-\De_0/\epsilon_0$. If we thus want to avoid the generation
of $\De$, $\lambda^e$, and $\lambda^y$ as defined in 
(\ref{TrunkHamiltonianUnShifted}), we set $\delta=0$, and the 
parameters of the generator have to satisfy the following conditions:
\begin{align}
  R\eta^{0,y}+\eta^y\lambda^z1_n=0\quad,\quad \eta^e=0\quad,\quad-\De
  \eta^z-R \eta^x+\eta^y\omega_0=0
\end{align}

Again the parameters of the generator are only defined up to a
common factor.  Choosing $\eta^e=0$, $\eta^x=-\omega_0 \lambda^xf/2$,
$\eta^z=-\omega_0\lambda^zf/2$ renders $\mathcal{O}_5$ zero and yields
$\eta^{0,y}=\lambda^x\lambda^zf1_n/2$ and $\eta^y=-R\lambda^xf/2$.
Thus all neglected operators but $\mathcal{O}_1$ and $\mathcal{O}_2$
are zero.  We obtain the following flow equations: 
\begin{align}
\label{Version3}
\dl R&=-R{\lambda^x}^2f1_n\quad,\quad \dl
E=-\omega_0({\lambda^x}^2+{\lambda^z}^2)f/2\\\notag \dl
\lambda^x&=-\omega_0^2\lambda^xf+R^2\lambda^xf
-{\lambda^z}^2\lambda^xf1_n\quad,\quad \dl
\lambda^z=-\omega_0^2\lambda^zf+{\lambda^x}^2\lambda^zf1_n
\end{align}

The set of equations in (\ref{Version3}) is equivalent to the set of
equations in (\ref{Version1d}). This can be seen by introducing
``new'' variables $\Delta'=\lambda^xR/\lambda'$,
$\epsilon'=\lambda^zR/\lambda'$ and 
${\lambda'}^2={\lambda^x}^2+{\lambda^z}^2$ and setting up their differential
equations, which coincide with (\ref{Version1d}). This demonstrates
that keeping the Hamiltonian form-invariant during the flow preserves
the unitary equivalence with respect to the initial Hamiltonian for
this special unitary transformation. 

If we want the initial Hamiltonian to remain form-invariant during the
flow after having shifted the bosonic mode by $\theta$, the constants
have to satisfy the following relations:
\begin{eqnarray}
\label{Brokenconditions}
&-\epsilon\eta^y+\eta^x\omega_0-\eta^{0,y}\lambda^z+2\eta^y\lambda^z\delta=0&\\
&-\De\eta^z-\epsilon\eta^x+\eta^y\omega_0+2\eta^x\lambda^z\delta=0&
\end{eqnarray}

After the shift, $\lambda^e$ is naturally generated which was not  
present in the previous schemes. In order to compare the flow
equations with the above versions, we have to couple the flow of
$\lambda^e$ with the flow of $\lambda^z$, i.e.
$\lambda^e=\theta\lambda^z$. This sets another condition on the
parameters of the generator, i.e.
$\eta^e=-\theta\De\eta^y/\omega_0+\theta\eta^z$.  If we further choose
$\eta^y=-\De\lambda^zf/2$ we obtain $\eta^z=-\omega_0\lambda^zf/2$,
$\eta^x=0$, $\eta^{0,y}=\epsilon\De f/2-\De\lambda^z\delta f$ and
$\eta^e=\theta\De^2\lambda^zf/ (2\omega_0)-\theta\omega_0\lambda^zf/2$
with the factor $f$ to be determined later.  With
$\bar{\epsilon}=\epsilon-\theta{\lambda^z}^2/\omega_0$, this yields
the following flow equations :
\begin{align}
\label{Forminv_Brok_Rabi}
\begin{split}
  \dl \De&=-\De{\lambda^z}^2f1_n +\De
  (\bar{\epsilon}+\lambda^z(\delta-\theta\lambda^z/\omega_0))^2f\quad,\quad
  \dl \bar{\epsilon}=-\De^2\bar{\epsilon} f
  +2\De^2\lambda^z(\delta-\theta\lambda^z/\omega_0)f\quad,\quad\\
  \dl \lambda^z&=\lambda^z(\De^2-\omega_0^2)f\quad,\quad \dl
  E=-\omega_0{\lambda^z}^2f/2+\theta^2(\De^2-\omega_0^2)
  {\lambda^z}^2f/(2\omega_0)
\end{split}
\end{align}
Recalling the initial condition of the energy shift
$E_0'=E_0+\theta^2\lambda_0^2/(4\omega_0)$ defined in Eq.
(\ref{InitialEnergyShift}) we see that the flow equations are
equivalent to the flow equations of Version $d$ if we set $f=1$ and
$\delta=\lambda^e/\omega_0=\theta\lambda^z/\omega_0$. This choice of
the $\ell$-dependent shift coincides with the expression 
(\ref{BosonicShift}) if we set $\LZR=0$.

This is a remarkable result. It shows that if one imposes invariance
of the flow equations with respect to unitarily equivalent initial
Hamiltonians and chooses the truncation scheme that leaves the
Hamiltonian form-invariant, the flow equations are uniquely
determined. It also shows that normal ordering with respect to the
$\ell$-dependent mode $\bb=b+\delta$ leads to reasonable results.

\begin{figure}[t]
  \begin{center}
    \epsfig{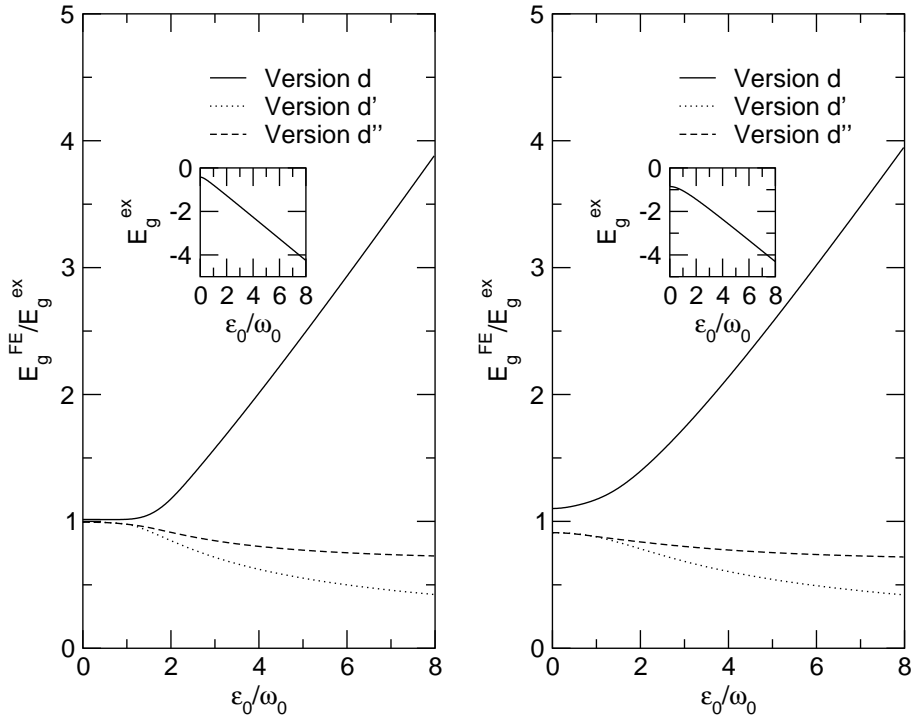}
    \caption{The ground-state energy $E_g^{FE}$ obtained from the 
      form-invariant flow for $\De_0/\omega_0=0.5$ (left hand side)
      and $\De_0/\omega_0=1.5$ with $\lambda_0/\omega_0=1$ as a
      function of the bias $\epsilon_0$ relative to the exact
      ground-state energy $E_g^{ex}$, shown in the panel. The primed
      versions are including the flow of the neglected operators
      $\mathcal{O}_1$ and $\mathcal{O}_2$ (see text).}
    \label{GSE_FI}
\end{center}
\end{figure}

We now want to check the quality of the form-invariant truncation scheme. 
In Figure \ref{GSE_FI} the ground-state energy
$E_g^{FE}$ obtained by the set of
equations (\ref{Version1d}) is shown relative to the exact
ground-state energy $E_g^{ex}$ for two different tunnel-matrix
elements $\Delta_0/\omega_0=0.5$ (left hand side) and
$\Delta_0/\omega_0=1.5$ (right hand side). Drastic deviations from the
exact result are seen in the regime $\epsilon_0/\omega_0\geq1$. This
means that the neglected operator $\mathcal{O}_1$ of Eq.
(\ref{Neglect_Rabi}) becomes relevant and has to be taken into
account.
 
In order to demonstrate that the flow equations can be improved
systematically, we will now consider higher order terms of the bosonic
operators in their normal ordered representation. For the normal ordering procedure see Appendix \ref{AppNormalOrdering}.
Since we set $\delta=0$ for all
$\ell$, normal ordering is defined with the respect to the unshifted
mode, i.e. $\bar{b}=b$.  Redefining $\mathcal{O}_1\equiv\sigma_x
\kappa_1:(b+b\dag)^2:$, the commutator $[\eta,\mathcal{O}_1]$
yields 
\begin{align} 
\label{Zusatz_Rabi}
[\eta,\mathcal{O}_1]&=2\sz\eta^{y,0}\kappa_1:(b+b\dag)^2:+2\sz\eta^y
\kappa_1(:(b+b\dag)^3:+2\langle(b+b\dag)^2\rangle:(b+b\dag):)\notag\\
&+2i\sy\eta^z\kappa_1:(b-b\dag)(b+b\dag)^2:\quad.
\end{align}
We first neglect the trilinear operators and the bilinear
operator of type $\mathcal{O}_2$ (see Eq. (\ref{Neglect_Rabi})). 
The extended flow equations then read ($f=1$)
\begin{align}   
\label{Version1d_p} 
\begin{split}   
  \dl \De&=-\De{\lambda^z}^21_n+\De\epsilon^2\quad,\quad
  \dl  \epsilon=-\epsilon\De^2\quad,\quad\dl \kappa_1=\De{\lambda^z}^2/2 \\
  \dl
  \lambda^z&=\lambda^z(\De^2-\omega_0^2)+4\lambda^z\De\kappa_11_n\quad,\quad
  \dl E=-\omega_0{\lambda^z}^2f/2\quad.
\end{split}
\end{align}
We will refer to this set of flow equations as Version $\text{d}'$.

To see if this improvement is systematic we will now include also the
corrections that come from the neglected operator of type
$\mathcal{O}_2$. Redefining
$\mathcal{O}_2\equiv\sigma_z\kappa_2:(b+b\dag)^2:$, we obtain similar
commutator relations for $[\eta,\mathcal{O}_2]$ as we got in 
(\ref{Zusatz_Rabi}):
\begin{align} 
\label{Zusatz_Rabi_Two}
[\eta,\mathcal{O}_2]&=-2\sx\eta^{y,0}\kappa_2:(b+b\dag)^2:-2\sx\eta^y
\kappa_2(:(b+b\dag)^3:+2\langle(b+b\dag)^2\rangle:(b+b\dag):)\notag\\
&-2i\sy\eta^z\kappa_2:(b-b\dag)(b+b\dag)^2:
\end{align}
The effect of including the operator $\mathcal{O}_2$ in the flow equations is the following: The conditions for the
constants of the generator that assure the form-invariance of the
Hamiltonian slightly change, see Eq. (\ref{syp_Rabi}).  The flow
equations thus read ($f=1$)
\begin{align}   
\label{Version1d_pp}    
\begin{split}   
  \dl
  \De&=-\De{\lambda^z}^21_n+\De\epsilon(\epsilon+4\kappa_2)\quad,\quad
  \dl  \epsilon=-(\epsilon+4\kappa_2)\De^2\\
  \dl
  \lambda^z&=\lambda^z(\De^2-\omega_0^2)+4\lambda^z\De\kappa_11_n\quad,\quad
  \dl  E=-\omega_0{\lambda^z}^2f/2\\
  \dl
  \kappa_1&=\De{\lambda^z}^2/2-\De(\epsilon+4\kappa_2)\kappa_2\quad,\quad\dl
  \kappa_2=\De(\epsilon+4\kappa_2)\kappa_1\quad.
\end{split}
\end{align}
We will refer to this set of flow equations as Version $\text{d}''$.

In Fig. \ref{GSE_FI} one sees that the extended flow equations yield
a systematic improvement ranging over the whole parameter space.
Nevertheless, the agreement with the exact result remains rather poor
for $\epsilon_0/\omega_0\geq1$. Only if one considers the
renormalization of the bath mode $\omega_0$, one obtains results
within a few percent relative error over the whole parameter range.
Regarding the spin-boson model, it is preferable to employ
the canonical generator since the bath modes remain unrenormalized in
the thermodynamic limit \cite{Sta02}.

\section{Flow of Observables}

We will now investigate the flow of observables. In order to
characterize the quality of the flow equations, normally sum rules are
derived expressing the fact that $\sigma_i^2=1$ or (anti-)commutation relations should hold for all $\ell$ with $i=x,y,z$ \cite{Keh97,Rag99}. 
As will be pointed out in the end of this
section, these sum rules can be misleading. We will therefore
contrast the expectation value $\LZR$ as it follows from the flow
equation approach with the numerically exact solution. Furthermore, we
will compare the flow equation results with the numerically exact
fixed point of the operator flow on the operator level. To do so, we
will give a unique decomposition of the fixed point operator into a
basis of normal ordered bosonic operators.

\subsection{Flow Equations for the Pauli Matrices}

In order to take advantage of the simple form of the fixed point Hamiltonian
when calculating expectation values of observables, the observable has to be subjected to the same sequence of unitary transformations as the Hamiltonian. 
The flow equations for the Pauli spin matrices thus read 
$\dl\sigma_i=[\eta,\sigma_i]$. Again the flow equations generate an infinite series of operators and one needs a suitable truncation and decoupling scheme. The $i$-component of the Pauli spin matrices as a
function of the flow parameter $\ell$ shall be given by
\begin{align}
\label{Trunc_SigmaXZ_Rabi}
  \begin{split}
    \sigma_i (\ell)&=g_i(\ell)\sigma_x  +h_i(\ell)\sigma_z+f_i(\ell)
    +\sx\chi^{x,i}(\ell)(b+b\dag)+i\sy\chi^{y,i}(\ell)(b-b\dag)
    +\sz\chi^{z,i}(\ell)(b+b\dag)\quad,
  \end{split}
\end{align}
with $i=x,z$. We want to emphasizes that the constant term $f_i$ is
indeed generated even though it seems to contradict the theorem of the
invariance of the trace under unitary transformations. A short
discussion is given in Appendix \ref{Aconst_f}.

The flow of the $y$-component of the Pauli spin matrices is given by
\begin{align}\label{Trunc_SigmaY_Rabi}
  i\sigma_y (\ell)=g_y(\ell)i\sigma_y
  +\sx\chi^{x,y}(b-b\dag)+\sz\chi^{z,y}(b-b\dag)\quad.
\end{align}  
These are the most general expansions up to linear bosonic operators
with real coefficients that can evolve from the Pauli spin matrices
under the flow equations, i.e. from $\sigma_i(\ell=0)=\sigma_i$.

The commutator $[\eta,\sigma_i]$ with $i=x,z$ yields the following
contributions:
\begin{align}
  [\eta^{0,y},\sigma_i(\ell)]&=2\sz g_i\eta^{0,y}-2\sx
  h_i\eta^{0,y}+2\sz \eta^{0,y}\chi^{x,i}(b+b\dag)\\\notag
  &-2\sx \eta^{0,y}\chi^{z,i}(b+b\dag)\\
  [\eta^e,\sigma_i(\ell)]&=2\sx\eta^e\chi^{x,i}+2\sz\eta^e\chi^{z,i}\\
  [\eta^x,\sigma_i(\ell)]&=-2i\sy h_i\eta^x(b-b\dag)+2\eta^x\chi^{x,i}
  -2\sz \eta^x\chi^{y,i}(b-b\dag)^2\\\notag
  &-i\sy \eta^x\chi^{z,i}\{(b-b\dag),(b+b\dag)\}\\
  [\eta^y,\sigma_i(\ell)]&=2\sz g_i\eta^y(b+b\dag)-2h_i\sx
  \eta^y(b+b\dag)\\\notag &+2\sz
  \eta^y\chi^{x,i}(b+b\dag)^2+2\eta^y\chi^{y,i}
  -2\sx \eta^y\chi^{z,i}(b+b\dag)^2\\
  [\eta^z,\sigma_i(\ell)]&=2i\sy g_i\eta^z(b-b\dag) +i\sy
  \eta^z\chi^{x,i}\{(b-b\dag),(b+b\dag)\}\\\notag &+2\sx
  \eta^z\chi^{y,i}(b-b\dag)^2+2\eta^z\chi^{z,i}
\end{align}
The commutator $[\eta,i\sy]$ is given by:
\begin{align}
  [\eta^{0,y},\sy(\ell)]&=2\sz \eta^{0,y}\chi^{x,y}(b-b\dag)
  -2\sx \eta^{0,y}\chi^{z,y}(b-b\dag)\\
  [\eta^x,\sy(\ell)]&=-2\sz g_y\eta^x(b-b\dag)-2i\sy \eta^x\chi^{z,y}
  (b-b\dag)^2\\
  [\eta^y,\sy(\ell)]&=\sz \eta^y\chi^{x,y}\{(b+b\dag),(b-b\dag)\}
  -\sx \eta^y\chi^{z,y}\{(b+b\dag),(b-b\dag)\}\\
  [\eta^z,\sy(\ell)]&=2\sx g_y\eta^z(b-b\dag) +2i\sy
  \eta^z\chi^{x,y}(b-b\dag)^2
\end{align}

Again, $\{.,.\}$ denotes the anti-commutator. To understand which operators
can transform into one another, we give a list of operators and their
behavior under parity transformation (P) and Hermitian
conjugation (H) ($x\equiv(b+b\dag)$, $p\equiv(b-b\dag)$) :\\
\begin{align}
  \notag \text{
\begin{tabular}{c||c|c|c|c|c|c|c|c|c|c|c|c}
&$1$&$\sx$&$i\sy$&$\sz$&$x$&$p$&$\sx x$&$\sx p$&$i\sy x$&$i\sy p$&$
\sz x$&$\sz p$\\
\hline
P&$+$&$+$&$-$&$-$&$+$&$-$&$-$&$-$&$+$&$+$&$+$&$+$\\
H&$+$&$+$&$-$&$+$&$+$&$-$&$+$&$-$&$-$&$+$&$+$&$-$
\end{tabular}}
\end{align}
\newline

In order to close the flow equations, we neglect normal ordered
bosonic bilinears where normal ordering is defined with respect to the
shifted bosonic mode $\bar{b}=b+\delta/2$. Thus, one obtains the
following set of linear differential equations for the $i$-component
of the Pauli spin matrices with $i=x,z$:
\begin{align}
  \dl g_i&=-2h_i
  \eta^{0,y}+2\eta^e\chi^{x,i}-2\eta^z\chi^{y,i}1_n-2\eta^y\chi^{z,i}1_n
  +2\eta^y\chi^{z,i}\delta^2\\
  \dl h_i&=2g_i
  \eta^{0,y}+2\eta^y\chi^{x,i}1_n+2\eta^x\chi^{y,i}1_n+2\eta^e\chi^{z,i}
  -2\eta^y\chi^{x,i}\delta^2\\
  \dl f_i&=2\eta^{x}\chi^{x,i}+2\eta^y\chi^{y,i}+2\eta^z\chi^{z,i}
\label{Brokenconst_f}\\
\dl \chi^{x,i}&=-2h_i\eta^y-2\eta^{0,y}\chi^{z,i}+4\eta^y\chi^{z,i}\delta\\
\dl \chi^{y,i}&=2g_i\eta^z-2h_i\eta^x+2\eta^x\chi^{z,i}\delta-2\eta^z\chi^{x,i}\delta\\
\dl
\chi^{z,i}&=2g_i\eta^y+2\eta^{0,y}\chi^{x,i}-4\eta^y\chi^{x,i}\delta
\end{align}
The flow equations for the $y$-component read:
\begin{align}
  \dl g_y&=-2\eta^z\chi^{x,y}1_n+2\eta^x\chi^{z,y}1_n\\
  \dl
  \chi^{x,y}&=2g_y\eta^z-2\eta^{0,y}\chi^{z,y}-2\eta^y\chi^{z,y}\delta
  \quad,\quad \dl
  \chi^{z,y}=-2g_y\eta^{0,y}x+2\eta^{0,y}\chi^{x,y}-2\eta^y\chi^{x,y}\delta
\end{align}
If no approximation was made, $\sigma_i^2(\ell)=1$ would hold for all
$\ell$ and $i=x,y,z$.  Taking the expectation value with respect to
the bilinear Hamiltonian of the shifted modes the relation should hold
approximately for $i=x,z$:
\begin{align}
\begin{split}
  \langle \sigma_i^2(\ell)\rangle &=g_i^2+h_i^2+f_i^2+
  (\chi^{x,i}\chi^{x,i}+\chi^{y,i}\chi^{y,i}+\chi^{z,i}\chi^{z,i})1_n\\
  &+2(g_i\langle\sigma_x \rangle+h_i\langle\sigma_z \rangle)f_i
  +2(\chi^{x,i}\langle\sz \rangle-\chi^{z,i}\langle\sx \rangle)\chi^{y,i}\\
  &+(\chi^{x,i}\chi^{x,i}+\chi^{z,i}\chi^{z,i})\delta^2
  -2((g+\LXR f)\chi^{x,i}+(h+\LZR f)\chi^{z,i})\delta\\
  &\approx1
\end{split}
\end{align}
For the $y$-component we obtain:
\begin{align}
  \langle \sigma_y^2(\ell)\rangle &=g_y^2+
  (\chi^{x,y}\chi^{x,y}+\chi^{z,y}\chi^{z,y})1_n\approx1
\end{align}
Other conservation relations follow e.g. from the commutator
$[\sx(\ell),\sz(\ell)]=-2i\sy(\ell)$.
These relations can be used to assess the validity and the quality of
the flow equations but they cannot assure whether the scheme will yield the correct results. We will comment on this point at the end of this section.

\subsection{Numerical Results for the Expectation Value of $\sigma_z$}

Measurable quantities other than the ground-state energy are determined
by means of the operator flow. In this subsection we will discuss the
expectation value $\LZR$ as it follows from the different versions of
the flow equation approach. The expression is given by
\begin{align}
  \LZR={}^*\langle\sigma_z(\ell=\infty)\rangle^*=
  g(\ell=\infty){}^*\langle\sigma_x\rangle^*+
  h(\ell=\infty){}^*\langle\sigma_z\rangle^*+f(\ell=\infty)\quad.
\end{align}
Here, ${}^*\langle...\rangle^*$ denotes the ground-state expectation
value with respect to the fixed point Hamiltonian $H(\ell=\infty)$.
\begin{figure}[t]
  \begin{center}
    \epsfig{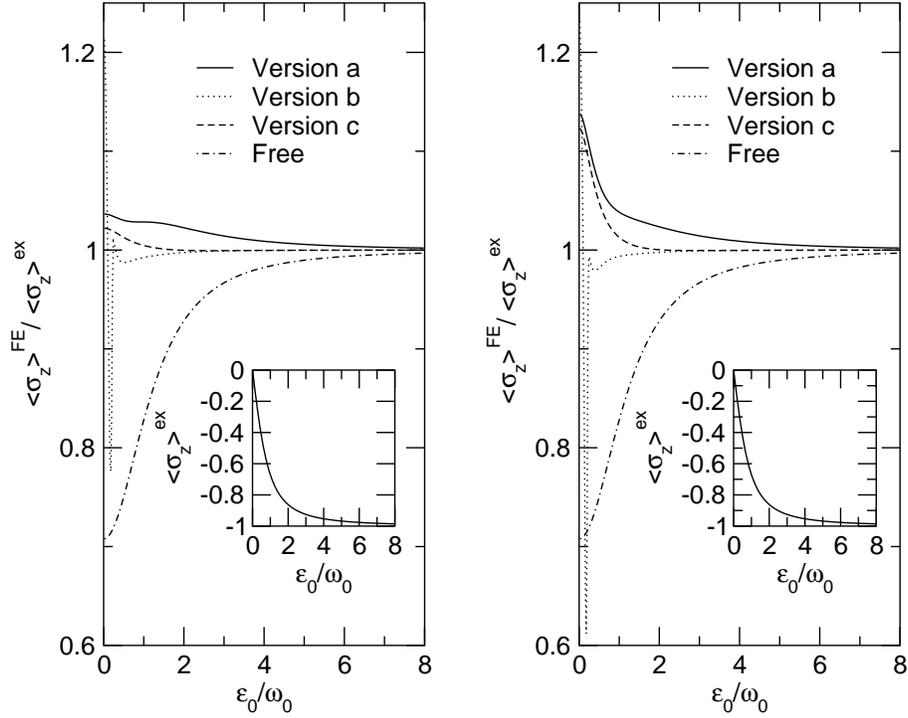}
    \caption{The expectation value $\langle\sz\rangle^{FE}$ obtained 
      by different canonical generators with $\psi=0$ and $\theta=0$
      for $\De_0/\omega_0=1.5$ with $\lambda_0/\omega_0=1$ as function
      of the bias $\epsilon_0$ relative to the exact expectation value
      $\langle\sz\rangle^{ex}$, shown in the panel. The expectation
      values for the bosonic shift
      $\delta=\sum_j\langle\sigma_j\rangle\lambda^j/\omega_0$ were
      evaluated directly according to Eqs. (\ref{EVdirectly}) (left
      hand side) and self-consistently according to Eqs.
      (\ref{SelfconsistingSet}) (left hand side).}
    \label{Abb4}    
\end{center}
\end{figure}
\begin{figure}[t]
  \begin{center}
    \epsfig{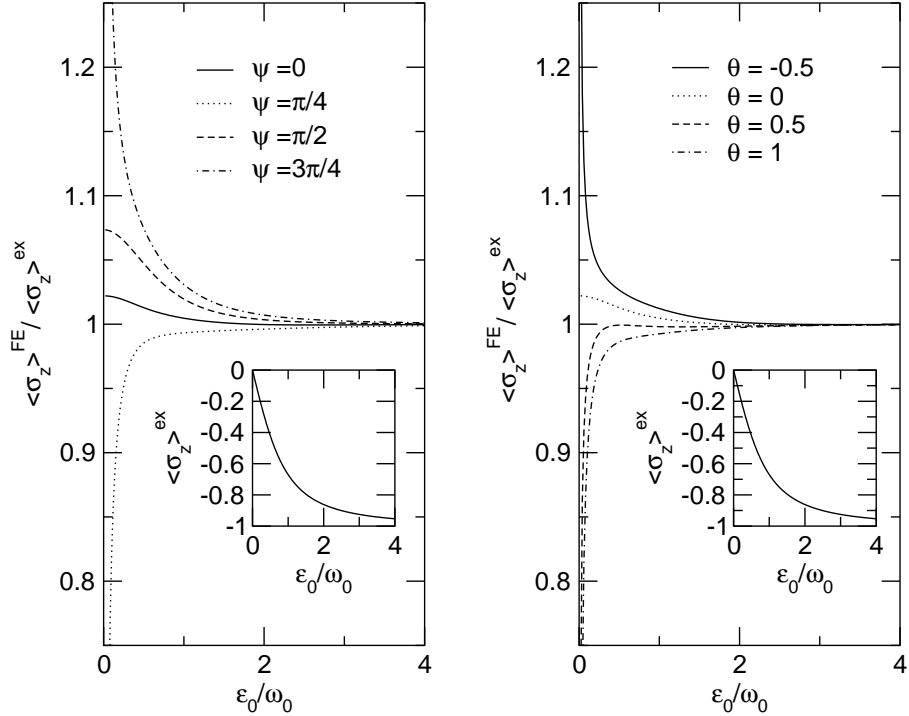}
    \caption{The expectation value $\langle\sz\rangle^{FE}$ 
      obtained by the canonical generator of Version c with
      $\delta=\sum_j\langle\sigma_j\rangle\lambda^j/\omega_0$ for
      $\De_0/\omega_0=1.5$ and $\lambda_0/\omega_0=1$ as a function of
      the bias $\epsilon_0$ relative to the exact expectation value
      $\langle\sz\rangle^{ex}$, shown in the panel. The parameters of
      the initial Hamiltonian are given by $\theta=0$ and various
      $\psi$ (left hand side) and $\psi=0$ and various $\theta$ (right
      hand side).}
    \label{Abb5c}
\end{center}
\end{figure}

In Fig. \ref{Abb4} we contrast the results for the different
generators which were discussed in the last section,
$\LZR^{FE}$, with the numerically exact solution $\LZR^{ex}$. We
choose $\psi=0$ and $\theta=0$ for the initial Hamiltonian and we will
employ the flow equations obtained by the generalized normal ordering
procedure, i.e.
$\delta=\sum_j\langle\sigma_j\rangle\lambda^j/\omega_0$.

On the left hand side of Fig. \ref{Abb4}, the expectation values in
the expression of $\delta$ are determined directly according to Eqs.
\ref{EVdirectly}. On the right hand side of Fig. \ref{Abb4}, the
expectation values are evaluated self-consistently according to Eqs.
\ref{SelfconsistingSet}.  For $\epsilon_0/\omega_0\geq1$, the best
results are obtained by the generator of Version b with the direct
evaluation of the expectation values. But deviations from the exact
solution in the region $\epsilon_0/\omega_0\leq1$ are significant. In
the latter region the generator of Version c yields the best results.
We recall that the ground-state energy was best approximated by the 
generator of Version b with the self-consistent evaluation of the 
expectation values entering the bosonic shift $\delta$. This demonstrates 
that the ``best'' generator and ``best'' procedures of taking account 
of the neglected terms might depend on the physical
quantity under consideration.

We will now also include the initial unitary transformation on the
two-dimensional spin-Hilbert space, label by $\psi$ and the initial
bosonic shift $\theta$ in our discussion.  We will use the generator
of Version c with the direct evaluation of the expectation values. On
the left hand side of Fig. \ref{Abb5c}, we vary $\psi$ with
$\theta=0$; on the right hand side of Fig. \ref{Abb5c} we vary
$\theta$ with $\psi=0$.

Regardless the initial Hamiltonian, the flow equation results differ
from the exact solution in the region $\epsilon_0/\omega_0\leq2$. But
some initial Hamiltonians provoke more significant deviations than
others. Good results over the whole parameter space are obtained by
combining non-zero values of $\psi$ and $\theta$ which
``compensate'' their errors, e.g. $\psi=\pi/32$ and $\theta=-0.2$. 
Nevertheless, we were not able to given an objective procedure 
how to choose the optimal initial Hamiltonian - {\it a priori}.

\subsection{Operator Fixed point}

It is possible to compare the exact results with the
flow equation approach not only on the spectral but also on the
operator level. For this we have to diagonalize the Hamiltonian in 
this basis in 
which the corresponding ``diagonal'' Hamiltonian $H_0$ of the flow
equation approach is diagonal. Let now 
$H_{\text{D}}=UHU\dag$ denote the diagonalized Hamiltonian, then
$\sigma_i^*=U\sigma_iU\dag$ is the operator to be compared with
$\sigma_i(\ell=\infty)$ stemming from the flow equation approach, with
$i=x,y,z$. To do so we will decompose $\sigma_i^*$ in a set of
operators which are created by the corresponding flow equations.

\begin{figure}[t]
  \begin{center}
    \epsfig{file=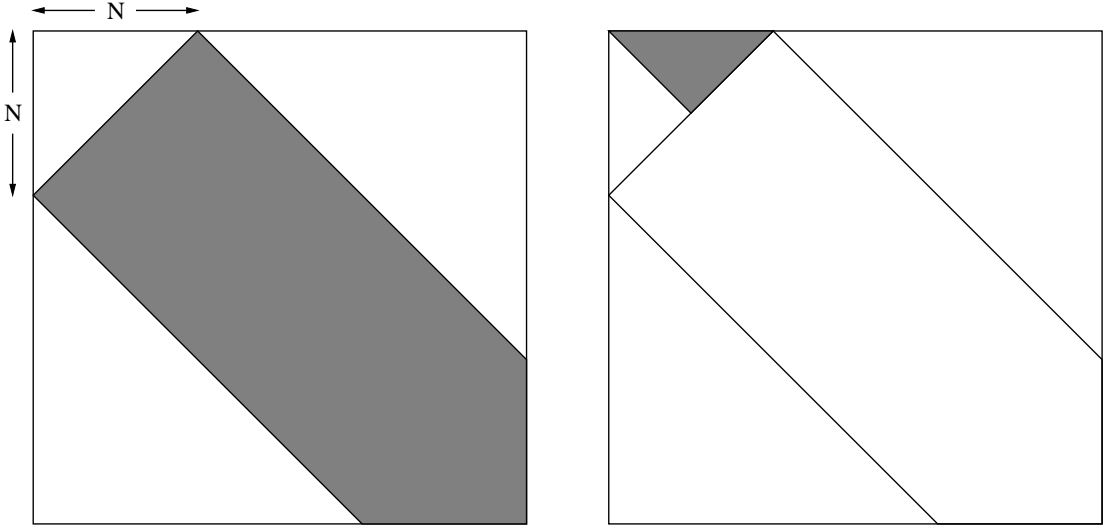,height=7cm,angle=0}
    \caption{Left hand side: Matrix structure of a normal ordered 
      bosonic operator consisting of $N$ bosonic operators with
      respect to the canonical basis (see text). The dark area
      indicates non-zero entries. Right hand side: The dark area
      indicates the matrix elements of an arbitrary matrix which are
      uniquely determined by normal ordered operators consisting of up
      to a certain number of bosonic operators (see text).}
    \label{MatrixStructure}
    \end{center}
\end{figure}

If one uses an expansion which is {\em normal ordered} in the bosonic
operators the decomposition can be obtained numerically without any
approximation \cite{error}. The reason for this is that the bosonic
ladder operators cannot compensate each other and then act on lower
bosonic subspaces. To make this more explicit the general matrix
structure of a normal ordered operator consisting of $N$ bosonic
operators is shown on the left hand side of Figure
\ref{MatrixStructure}, taking the set $\{|\nu\rangle\}$ as basis with
$|\nu\rangle\equiv(b\dag)^{\nu}/\sqrt{\nu!}|0\rangle$ and
$b|0\rangle=0$, $\nu$ being a positive integer. The dark area contains
non-zero entries whereas the white area contains no entries. In case
of a non-normal ordered operator the white, upper left triangle would
also contain non-zero entries.
 
As an explicit choice of the operator basis for real symmetric
operators like $\sigma_x$ and $\sigma_z$ we choose the set
$\{o:(b+b\dag)^n(b-b\dag)^{2m}:,
o':(b+b\dag)^{n'}(b-b\dag)^{2m'+1}:\}$, where $o=1,\sx,\sz$ and
$o'=i\sigma_y$. The operator basis for real antisymmetric operators is
obtained by interchanging $o$ and $o'$. In the following we will only
consider the flow of real symmetric operators.  The results also hold
for the real antisymmetric case.

We want to decompose a real symmetric operator into a set of finite
operators.  Considering all operators of the basis given above with
less or equal than $2N$-bosonic operators, we obtain a finite basis of
$3\sum_{m=0}^{N}\sum_{n=0}^{2(N-m)}+
\sum_{m'=0}^{N}\sum_{n'=1}^{2(N-m')}=(N+1)(4N+3)$ operators. Summing
up the independent matrix elements which are uniquely determined by
the normal ordered operators containing up to $2N$ bosonic modes, we
obtain $\sum_{n=0}^{N}2(4n+1)+1=4(N+1)N+3(N+1)=(N+1)(4N+3)$. These
independent matrix elements are located at the upper left triangle of
the matrix, indicated as dark area on the right hand side of Figure
\ref{MatrixStructure}.

In order to complete the discussion we also consider all operators
with less or equal than $(2N+1)$-bosonic operators.  We then obtain a
basis with $3\sum_{m=0}^{N}\sum_{n=0}^{2(N-m)+1}+
\sum_{m'=0}^{N}\sum_{n'=0}^{2(N-m')}=(N+1)(4N+7)$ operators. Summing
up the independent matrix elements which are uniquely determined by
the normal ordered operators containing up to $2N+1$ bosonic modes, we
obtain $\sum_{n=0}^{N}2(4n+3)+1=4(N+1)N+7(N+1)=(N+1)(4N+7)$.

We thus obtain the same number of independent matrix elements and
basis ``vectors''. This confirms that our basis is complete and
linearly independent as we take $N\to\infty$.  Secondly, this shows
that the first $(N+1)(4N+3)$ coordinates of a real symmetric operator
with respect to a finite basis of operators up to $2N$ bosonic
operators are left unchanged if one goes over to a finite basis
including $2N+M$ bosonic operators ($M>0$).

We can thus {\em exactly} determine the coefficients of our basis up
to any number of bosonic excitations $N$ which $\sigma_i^*$ is
composed of. This shows that choosing a set of normal ordered bosonic
operators as a basis yields a systematic approximation of any
operator. If one is only interested in the system dynamics at low
energies it thus suffices to consider
only up to $N$ bosonic operators with $N=2$ say.

To determine the coefficients numerically one has to work with a
specific basis. Up to now we have only specified the basis of the
bosonic Hilbert space. Choosing $H_0=-\frac{\De_0}{2}\sigma_x +
\omega_0 b\dag b$ to be diagonal we are led to the basis
$\{|e,\nu\rangle\}$ with the first quantum number $e=0,1$ denoting the
eigenstates of $\sigma_x$ and the second quantum number denoting the
eigenstates of $b\dag b$.
Choosing $H_0=\frac{\epsilon}{2}\sigma_z + \omega_0 b\dag b$  
(Version $1b$) or the diagonalized representation
$H_0=\frac{R}{2}\sigma_z + \omega_0 b\dag b$ of Version $1c$, we would choose
the first quantum number $e=0,1$ to denote the eigenstates of
$\sigma_z$.
   
Considering all operators with less or equal than $2N$-bosonic
operators, we end up to solve a linear equation $Ax=b$, with $A$ being
a quadratic matrix and $x,b$ being vectors with dimensions
$(N+1)(4N+3)$.  The coefficients of the matrix $A$ are obtained by the
following matrix representations of normal ordered bosonic operators:
\begin{align}
  \notag &\langle e,\mu|o:(b+b\dag)^n(b-b\dag)^{2m}:|e',\nu\rangle
  =\langle e|o|e'\rangle
  \sum_{k=0}^n{n\choose k}\sum_{l=0}^{2m}{2m \choose l}(-1)^{2m-l}\\
  &\times\Theta(\nu-k-l)\sqrt{\frac{\mu!}{(N-k-l)!}}\sqrt{\frac{\nu!}{(N-k-l)!}}
  \delta_{\mu,\nu+n+2(m-k-l)}
\end{align}
The vector $b$ on the right hand side of the linear equation is given
by the $(N+1)(4N+3)$ independent matrix elements, located at the dark
area of the matrix of the right hand side of Figure
\ref{MatrixStructure}.

\subsection{Higher Orders}
In the expansion of the Pauli spin matrices of the last section we
have neglected all generated operators with more than one bosonic
operator.  In order to confirm that the expansion of the Pauli spin
matrices in normal ordered bosonic operators is indeed systematic we
will now upgrade our expansion and also include:
\begin{itemize}
\item{all generated operators up to two normal ordered bosonic
    operators}
\item{all generated operators up to three normal ordered bosonic
    operators}
\end{itemize}
In the following, normal ordering shall be defined with respect to the
bilinear Hamiltonian of the un-shifted mode, i.e. $\delta=0$. This
will simplify matters considerably. Choosing the parameters of the
initial Hamiltonian such that $\delta=0$ for all $\ell$, we are still
consistent within our normal ordering procedure.

The first extension, $\sigma_z^{new,2}$, includes the following terms,
where we introduce the abbreviations $x\equiv b+b\dag$ and $p\equiv
b-b\dag$ and where we also confine ourself to the discussion of $\sz$
in order to drop one index:
\begin{align}
  \notag \sigma_z^{new,2}&=\chi^1x+\sigma_x\psi^{x,+}:x^2:+
  i\sigma_y\psi^{y,+}:xp:\\
  &+\sigma_z\psi^{z,+}:x^2:+\sigma_x\psi^{x,-}:p^2:
  +\sigma_z\psi^{z,-}:p^2:
\end{align}
The second extension, $\sigma_z^{new,3}$, consists of the following terms:
\begin{align}
  \notag \sigma_z^{new,3}&=\psi^{1,+}:x^2:+\psi^{1,-}:p^2:
  +\sigma_x\varphi^{x,+}:x^3:+i\sigma_y\varphi^{y,+}:x^2p:\\
  &+\sigma_z\varphi^{z,+}:x^3:+\sigma_x\varphi^{x,-}:xp^2:+i\sigma_y
  \varphi^{y,-}:p^3:+\sigma_z\psi^{z,-}:xp^2:
\end{align}
The resulting flow equations for the upgraded truncation schemes are presented in Appendix \ref{HigherOrders}.

\subsection{Numerical Results} 
\begin{figure}[t]
  \begin{center}
    \epsfig{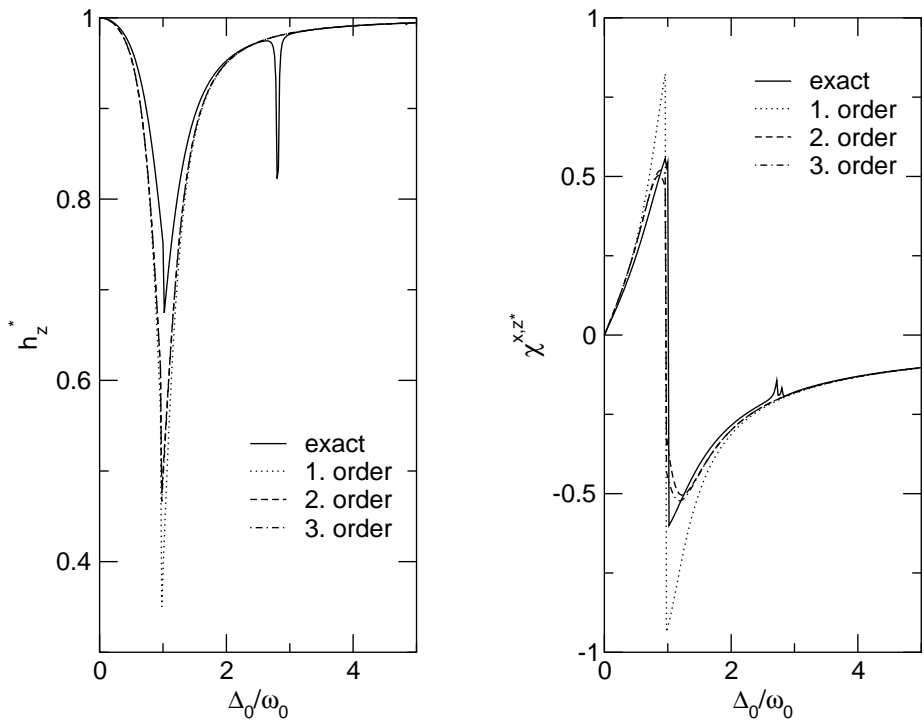}
    \caption{The fixed point parameters $h_z^*\equiv h_z(\ell=\infty)$ 
      (left hand side) and ${\chi^{x,z}}^*\equiv
      \chi^{x,z}(\ell=\infty)$ stemming from the symmetric flow
      equations of Version a for $\psi=0$, $\theta=0$,
      $\lambda_0/\omega_0=0.5$ and $\epsilon_0=0$ for different orders
      of truncation of the operator flow as a function of $\Delta_0$.
      The solid lines resembles the exact result.}
    \label{EpsilonZero_Rabi}
\end{center}
\end{figure}
\begin{figure}[t]
  \begin{center}
    \epsfig{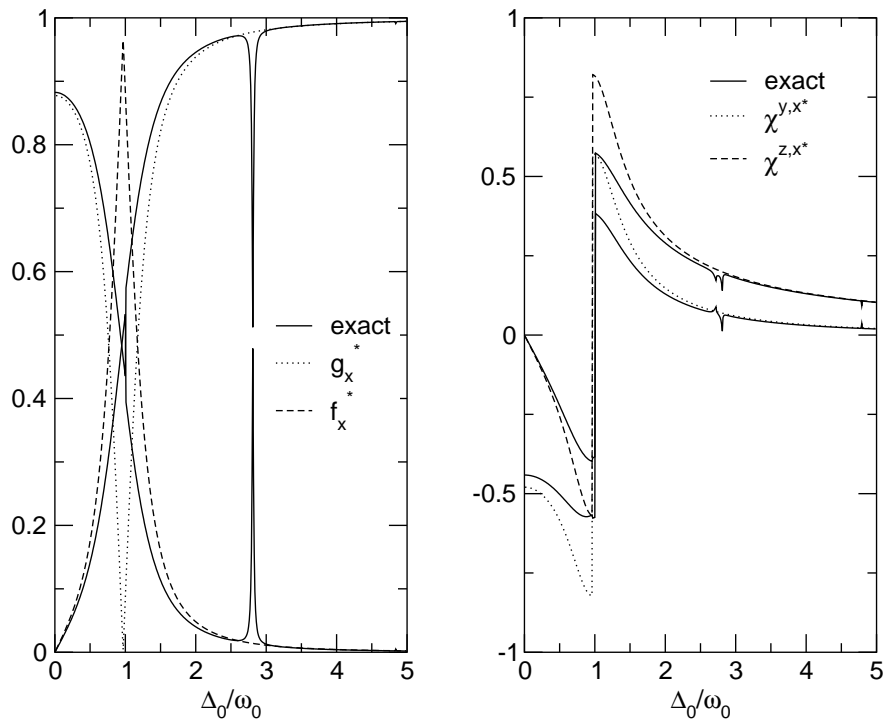}
    \caption{The parameters $g_x^*\equiv g_x(\ell=\infty)$ and 
      $f_x^*\equiv f_x(\ell=\infty)$ (left hand side) as well as
      ${\chi^{y,x}}^*\equiv \chi^{y,x}(\ell=\infty)$ and
      ${\chi^{z,x}}^*\equiv \chi^{z,x}(\ell=\infty)$ stemming from the
      symmetric flow equations of Version a for $\psi=0$, $\theta=0$,
      $\lambda_0/\omega_0=0.5$ and $\epsilon_0=0$ as function of
      $\Delta_0$. The solid lines resemble the analytic results.}
    \label{EpsilonZero_X_Rabi}
\end{center}
\end{figure}
\begin{figure}[t]
  \begin{center}
    \epsfig{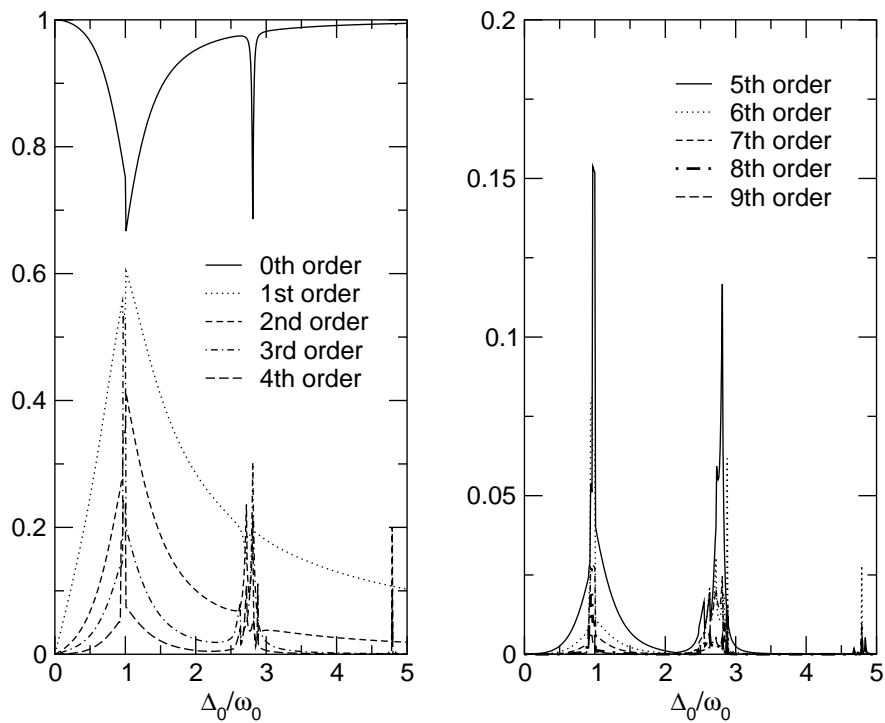}
    \caption{The sum of the absolute values of the coefficients 
      of all operators that consist of $n$ bosonic operators ($n$th
      order) which compose $\sigma_z^*$ for $\lambda_0/\omega_0=0.5$
      and $\epsilon_0=0$ as function of $\Delta_0$.}
    \label{Moments}
\end{center}
\end{figure}

We are now set to compare the fixed points of the operator flow
obtained from the flow equation approach with the exact results. We
can also see from the exact solution if the expansion into normal
ordered bosonic operators is preferable.

It turns out that the expansion into normal ordered operators is not without 
obstacles. Especially when the reflection symmetry is broken, i.e.
$\epsilon_0\neq0$, the final values of the coefficients delicately
depend on the initial parameters of the Hamiltonian. The reason for
this is that the unperturbed states cross when the interaction is
switched on and this effects the representation of the operator.  The
effect is enhanced by explicitly breaking certain symmetries.

Also the comparison of the operator flow
with respect to the different versions of the flow equations,
discussed in the previous section, is troublesome. 
Since the non-trivial versions for $\epsilon=0$ are
based on different diagonal Hamiltonians $H_0$, a direct comparison of
the fixed point parameters is not obvious.   

We therefore limit our investigations to the parameter regime where the
reflection symmetry is not broken, i.e. $\epsilon_0=0$. If we choose
the generator of Version a with $\psi=0$ and $\theta=0$, $\delta=0$
for all $\ell$, and if we consider the flow of the $z$-component of
the Pauli spin matrices, only two parameters $h_z$ and $\chi^{x,z}$
are being renormalized.  The final values $h_z^*\equiv
h_z(\ell=\infty)$ and ${\chi^{x,z}}^*\equiv\chi^{x,z}(\ell=\infty)$
are shown for the initial condition $\lambda_0/\omega_0=0.5$ in Figure
\ref{EpsilonZero_Rabi}, together with the results where we also
included the flow of bilinear ($2$. order) and trilinear ($3$. order)
bosonic operators, governed by Eqs. (\ref{gz_Zwei}) -
(\ref{psiz_Zwei}) and Eqs. (\ref{chix_Drei}) - (\ref{phiz_Drei}).

The fixed point coefficients $h_z^*$ and ${\chi^{x,z}}^*$ agree with
the exact solution unless the initial tunnel-matrix element $\Delta_0$
is close to a resonance, i.e. $\Delta_0\approx\omega_0$ or
$\Delta_0\approx3\omega_0$ \cite{Reso}. The spike at $\Delta_0\approx3\omega_0$
cannot be accounted for by any of the solutions obtained via flow
equations. But there is a significant improvement from the second
order to the first order result close to the resonance at
$\Delta_0\approx\omega_0$ especially in the case of ${\chi^{x,z}}^*$.
The improvement from third to second order in the case of ${\chi^{x,z}}^*$ is not as strong and the one particle parameter $h_z^*$ is almost left unchanged.

In Figure \ref{EpsilonZero_X_Rabi}, the results for the fixed point
operator $\sigma_x^*$ are shown as they follow from the flow equations
of Version a with the initial conditions $\lambda_0/\omega_0=0.5$ and
$\epsilon_0=0$. Four parameters $g_x^*\equiv g_x(\ell=\infty)$,
$f_x^*\equiv f_x(\ell=\infty)$,
${\chi^{y,x}}^*\equiv\chi^{y,x}(\ell=\infty)$, and
${\chi^{z,x}}^*\equiv\chi^{z,x}(\ell=\infty)$ are generated during the
flow. They show the same deficiencies with respect to the exact
solution as the results of Figure \ref{EpsilonZero_Rabi}. We want to
mention that the constant term $f_x^*$ is indeed generated, as can be
seen from the exact expansion.

To investigate the reason for the above discrepancies close to
resonances further, we are going to employ the numerically exact
solution and determine the expansion of the final operator
$\sigma_z^*=U\sigma_z U\dag$ including up to nine bosonic operators.
Instead of analyzing the graphs of all $115$ coefficients, we will
consider the sum of the absolute values of the coefficients that
belong to the operator class which consists of $n$ bosonic operators
($n$th-order).

The resulting nine graphs are shown in Figure \ref{Moments}. As can be
seen, the second order still contributes to the fixed point operator
considerably. Close to resonances even higher orders become important
for the operator expansion. This explains why the fixed point
parameter $h_z^*$ is not sufficiently recovered by the flow equation
approach even after including all terms up to three bosonic operators
into the flow equations.

In Appendix \ref{RabiPertubation}, the spikes of Figure \ref{Moments}
are related to degeneracies. The formalism thus breaks down at these
parameter configurations. This is related to the problem that occurs
when diagonalizing the Hamiltonian which is, strictly speaking, also
only possible for non-degenerate states.

Let us finally comment on sum rules that stem from operator relations
which remain invariant under unitary transformations.  Taking the
initial values for the Hamiltonian as in Figure
\ref{EpsilonZero_Rabi}, the flow equations of Version a yield the
exact sum rule $\langle\sz^2\rangle=h^2+(\chi^{x,z})^2=1$ at $T=0$ for
all $\ell$ and independent of the initial tunnel-matrix element
$\Delta_0$. The sum rule is thus not sensitive to the deviations
between the flow equation results and the exact solution, which become
especially drastic close to resonances, see Fig.
\ref{EpsilonZero_Rabi}. We observe the situation that two errors are
being canceled to yield the desired result. We therefore conclude that
the sum rule cannot be a sufficient criterion for the quality of the
operator flow. On the other hand, one cannot expect that the flow equations yield good results on all energy scales. Properties at low energies like the ground-state expectation value of $\sigma_z$ shown in Figs. \ref{Abb4} and \ref{Abb5c} still can be calculated with high precision. The typical deviations at resonances in the operator flow are averaged out.

\section{Conclusions}

This work addresses general questions concerning the flow equation approach 
such as optimization of the final results or invariance with respect
to the initial Hamiltonian, based on a simple non-trivial model.
The model is structurally similar to quantum impurity models and 
since the ``bath'' only consists of one mode, 
it is numerically exactly solvable. We
intended to demonstrate that a systematic improvement of the flow
equation approach is possible. In order to improve the flow equations
one can basically precede according to the following lines:
\begin{enumerate}
\item Most obviously, one can include more interaction terms in the truncation scheme of Hamiltonian and operator. This was
done for the Hamiltonian flow when employing the form-invariant
truncation scheme and a systematic improvement was seen. We did not
extend the truncation scheme for the canonical generator
because it is in principle not feasible for more realistic models with an arbitrary number of bosonic modes.
For the operator flow, the truncation scheme was extended up to
third order for a special parameter regime and the results were compared
with the exact solution on the operator level. Close to resonances,
the flow equation results showed significant deviations with respect
to the exact solution . These deviations were present even in the
upgraded truncation schemes since high orders of up to nine bosonic
operators still carried considerable weight. This is connected to the
general problem that the flow equation approach breaks down close to
degenerate states.
\item 
Another way to improve the flow equations is to consider the
neglected operators more thoroughly, 
i.e. to introduce a refined decoupling scheme. 
This was done by introducing a
$\ell$-dependent bosonic shift $\delta$ and neglecting normal ordered bilinear
bosonic operators with respect to this shifted mode. The bosonic shift was
deduced by formally diagonalizing the truncated Hamiltonian and than
decoupling the ``system'' from the ``bath''. The decoupling process
was not unambiguous and two different approaches were investigated.
These were labeled as direct and self-consistent evaluation of the 
system expectation values. 
The self-consistent approach turned out to yield better
results on the level of the Hamiltonian flow, the direct approach was
preferable on the level of the operator flow.
\item
A third possibility to obtain better results 
is to choose a different basis which the flow is defined on. 
As an example we want to mention the vertex flow introduced by Kehrein
\cite{Keh00}. We investigated
the operator flow with respect to the distinguished bosonic mode
$\bb=b+\delta/2$. The infinitesimal unitary transformations are then 
equivalent to an active {\em and} passive
transformation since the coefficients as well as the operator basis
$\bb$ are changing during the flow. But the numerical results turned out to be
worse than the ones based on the flow with respect to the unshifted mode $b$. We therefore did not include them in the discussion of the present paper. We want to emphasis, though, that there remains the possibility to improve the flow equation results along these lines.
\end{enumerate}

It is also pointed out that flow equations are, in general, 
not invariant with
respect to the initial Hamiltonian even though the Hamiltonians only
differ by a unitary transformation. We concluded that differences
are, in general, small and if one chooses a form-invariant truncation
scheme, the flow equations might not differ at all. But the fact that the results depend on the unitary representation of the initial Hamiltonian opens up the possibility to optimize the results by introducing an (arbitrary) number of parameters associated with possible unitary transformations and choosing them such that certain sum rules are fulfilled best. This strategy has been applied to the spin-boson model with external bias, where one parameter - associated with the shift of the bosonic operators - was chosen such that the sum rule of $\sigma_z$ was optimal for all $\ell$.\cite{Sta02}   
What had to be left open was how to choose the optimal initial Hamiltonian for the evaluation of a specific quantity - {\it a priori}.

The last part of the paper is dedicated to a detailed analysis of the operator flow. Since the flow equations are usually designed such that the Hamiltonian is diagonalized best, i.e. that the flow only involves few flow parameters, the transformation of the observable is more susceptible to uncontrolled approximations, i.e. higher order interaction terms are often neglected merely because they cannot be kept track of. For this reason, the exact operator fixed point was evaluated, represented in the basis which was determined by the specific choice of the generator. It turned out that the flow equations of the operator should include up to 115 interaction terms in order to adequately coincide with the exact operator fixed point on all energy scales. We also pointed out that exact sum rules resulting from the flow equations are mostly due to high symmetries of the operator flow, 
i.e. when only few terms are being generated. 
The assumption that the flow is well approximated if a
sum rule holds can thus be misleading as was shown in the last section. 
Nevertheless, the deviations at points of degeneracies of the operator flow with respect to the exact solution are unimportant for the low energy properties of the system. This was demonstrated by evaluating the ground-state expectation value of $\sigma_z$ within the most simple, but non-trivial truncation scheme.\\

We wish to thank F. Wegner for valuable discussions. This work was partially supported by the Deutsche Forschungsgemeinschaft.


\appendix

\section{The Independent Boson Model}
\label{IndependentBoson}
We want to give a brief introduction to the flow equation method based
on the exactly solvable Independent Boson Model. The Hamiltonian of
this model is given by
\begin{align}
\label{IndependentBosonH}
H&=H_0+V=\omega b^{\dagger}b+\epsilon c^{\dagger}c +\lambda
c^{\dagger}c(b+b\dag)\quad.
\end{align}
The $b^{(\dagger)}$ resemble bosonic, the $c^{(\dagger)}$ fermionic
operators. They obey the canonical commutation and anti-commutation
relations respectively.  The model can account for some relaxation
phenomena and is extensively discussed in the textbook by Mahan
\cite{Mah90}.

We set $\epsilon=\lambda^2/\omega$. Then the Hamiltonian of Eq.
(\ref{IndependentBosonH}) is unitarily equivalent to $H=\omega
b^{\dagger}b+ \sigma_z\lambda(b+b\dag)$, where $\sigma_z$ denotes the
$z$-component
of the Pauli spin matrices. This is the Rabi Hamiltonian (\ref{RabiHamiltonian}) with $\De_0=0$.\\

The model is easily solved by the unitary transformation
\begin{align}
\label{unitarytransformation}
U=\exp(-c^{\dagger}c\frac{\lambda}{\omega}(b-b\dag))
\end{align}
and we obtain the diagonalized Hamiltonian $UHU\dag=\omega
b^{\dagger}b$.

But we want to perform this unitary transformation continuously by
introducing a flow parameter $\ell$ and a family of unitarily
equivalent Hamiltonians $H(\ell)=U(\ell)HU\dag(\ell)$. We also want to
look closely at the transformed operator $c(\ell)\equiv
U(\ell)cU\dag(\ell)$ and question if an expansion of the operator in a
series of unbounded operators, namely $(b-b\dag)^n$, is well-defined.

The unitary operators $U(\ell)$ shall be defined by the generator
$\eta$ which governs the differential form of a continuous unitary
transformations as follows: $\dl H=[\eta,H]$.  A good choice for the
generator has proven to be $\eta=[H_0,V]$, which is likely to
eliminate the interaction in the limit $\ell\rightarrow\infty$ \cite{Weg94}. 
The $\ell$-dependent unitary operator $U(\ell)$ is related to the
generator $\eta$ through the differential equation $\dl U=\eta U$
which can be formally integrated to yield 
$U(\ell)=\mathcal{L}\exp(\int_0^{\ell}
d\ell'\eta(\ell'))$. The operator $\mathcal{L}$ denotes the $\ell$-ordering
operator, defined in the same way as the more familiar time-ordering
operator $T$. In fact, the differential form of the flow equations has
got the same structure as the Heisenberg equation of motion, but
complete formal equivalence is only achieved for explicitly
time-dependent Hamiltonians, since the generator $\eta$ is explicitly
$\ell$-dependent.\\

For the independent boson model the canonical generator reads
$\eta=-\omega\lambda c^{\dagger}c(b-b\dag)$ and we readily obtain
\begin{align}
  [\eta,H]=-\omega^2\lambda c^{\dagger}c(b+b\dag)
  -2\omega\lambda^2c^{\dagger}c\quad.
\end{align}
The following flow equations
\begin{align}
  \partial_{\ell}\lambda=-\omega^2\lambda\quad,\quad
  \partial_{\ell}\epsilon=-2\omega\lambda^2
\end{align}
are integrated to yield $\lambda(\ell)=\lambda\exp(-\omega^2\ell)$ and
$\epsilon(\ell)=\frac{\lambda^2}{\omega}\exp(-2\omega^2\ell)$.  Since
$[\eta(\ell),\eta(\ell')]=0$, the $\ell$-ordering operator $L$ becomes
trivial and we obtain for the $\ell$-dependent unitary operator
\begin{align}
\label{elldependentTransformation_IndBos}
U(\ell)=\exp(-c^{\dagger}c\frac{\lambda}{\omega}
(1-e^{-\omega^2\ell})(b-b\dag))\quad.
\end{align}
From Eq. (\ref{elldependentTransformation_IndBos}) we can obtain the
unique unitary operator for $\ell\to\infty$ which diagonalizes $H$ and
which was already given in Eq. (\ref{unitarytransformation}).

Given $U(\ell)$ one can determine the flow of the operator $c(\ell)$
directly:
\begin{align}
\label{non-normal}
c(\ell)&=U(\ell)cU\dag(\ell)=c
\exp(\frac{\delta\lambda(\ell)}{\omega}(b-b\dag))\\
&=c\exp(-\frac{1}{2}
\left(\frac{\delta\lambda(\ell)}{\omega}\right)^2)
\exp(-\frac{\delta\lambda(\ell)}{\omega}b\dag)
\exp(\frac{\delta\lambda(\ell)}{\omega}b)\\\label{normal} &\equiv
c\exp(-\frac{1}{2} \left(\frac{\delta\lambda(\ell)}{\omega}\right)^2)
\sum_{n=0}\left(\frac{\delta\lambda(\ell)}{\omega}\right)^n
\frac{:(b-b\dag)^n:}{n!}\quad,
\end{align}
where we introduced $\delta\lambda(\ell)=\lambda(1-e^{-\omega^2\ell})$
and defined normal ordering, denoted by $:...:$, by writing the
creation operator left from the annihilation operator. This
  definition of normal ordering resembles a special case of the
  general definition given in Appendix A and is valid at $T=0$. But
  from now on the general definition will be used.

We now apply the continuous transformation to the operator $c$ using
the differential form $\dl c=[\eta,c]$. The flow equations generate
the infinite series $c(\ell)=c\sum_{n=0}\gamma_n(\ell)(b-b\dag)^n$
with $\dl \gamma_{n+1}=\omega\lambda(\ell)\gamma_n$. Together with the
initial condition $\gamma_0=1$, $\gamma_n=0$ for $n\geq1$, this set of
differential equations can be solved to yield $\gamma_n=\frac{1}{n!}
(\frac{\delta\lambda(\ell)}{\omega})^n$. The flow equation result thus
coincides with the non-normal ordered form of $c(\ell)$ in Eq.
(\ref{non-normal}) if one expands the exponential function into a
Taylor-series.

At first sight there is no distinguished expansion of $c(\ell)$ in
bosonic operators since its generation depends on $\eta$. In order to
discuss a different scheme, we now define $c(\ell)$ by a series of
normal ordered operators, i.e.
$c(\ell)=c\sum_{n=0}\gamma_n(\ell){:(b-b\dag)^n:}$. We obtain the
following flow equations
\begin{align}
\label{diffequ_norm_ord}
\dl \gamma_{n+1}=\omega\lambda(\ell)(\gamma_n-(n+2)\gamma_{n+2})
\quad,
\end{align}
where we used the formula (see Appendix \ref{AppNormalOrdering})
\begin{align}
  (b-b\dag):(b-b\dag)^n:=:(b-b\dag)^{n+1}:+n\langle(b-b\dag)^2\rangle
  :(b-b\dag)^{n-1}:
\end{align}
at $T=0$, i.e. $\langle(b-b\dag)^2\rangle=-1$ with $\langle...\rangle$
denoting the bosonic ground-state expectation value.  Taking the same
initial conditions as in the case of the non-normal ordered expansion,
we see that the normal ordered expansion in Eq.  (\ref{normal}) solves
the set of differential equations (\ref{diffequ_norm_ord}), i.e.
$\gamma_n=\exp(-\frac{1}{2}
(\delta\lambda(\ell)/\omega)^2)\frac{1}{n!}
(\frac{\delta\lambda(\ell)}{\omega})^n$.

This is a remarkable result. Whereas the non-normal ordered expansion
of $c(\ell)$ reproduces the perturbative result in the coupling
$\delta\lambda$ for each coefficient $\gamma_n$, the normal ordered
expansion yields coefficients $\gamma_n$, which contain all powers of
$\delta\lambda$. Especially in view of later approximations, the
normal ordered version will then be more preferable, since it is
likely to go beyond a perturbative description.\\

After having recovered the correct flow of the observable via the flow
equation approach, we would like to investigate the ``stability'' of
the infinite expansion of $c(\ell)$ in unbounded operators.  For this
purpose, we consider the Green function $G(t)=-i\langle
Tc(t)c\dag\rangle$ and the spectral function
$A(\tilde{\omega})=-\text{Im} G(\tilde{\omega})/\pi$ with the time
ordering operator $T$, the Fourier transform $G(\tilde{\omega}) =\int
dte^{i\tilde{\omega} t}G(t)$ and $\langle\dots\rangle$ denoting the
ground-state expectation value with respect to $H$.  With
$\tilde{\lambda}\equiv\lambda/\omega$ we obtain \cite{Mah90}
\begin{align}
\label{Green_Function_InB}
G(t)&=-i\Theta(t)\exp(-\tilde{\lambda}^2
(1-e^{-i\omega t}))\quad,\\
A(\tilde{\omega})&=e^{-\tilde{\lambda}^2}
\sum_{n=0}\tilde{\lambda}^{2n}\frac{1}{n!}
\delta(\tilde{\omega}-n\omega)\quad.
\end{align}
The spectral function $A(\tilde{\omega})$ thus exhibits the polaronic
shift $\epsilon_p=-\lambda^2/\omega$ for $n=0$ and an equidistant
satellite structure separated by the oscillator frequency $\omega$
with exponentially decreasing weight.

Using flow equations, the Green function is best expressed as
\begin{align}
  G(t)=-i\Theta(t)\langle
  e^{iH(\ell=\infty)t}c\dag(\ell=\infty)e^{-iH(\ell=\infty)t}
  c(\ell=\infty)\rangle\quad,
\end{align}
because then the time evolution of the fermionic and bosonic operator
is that of free ones.

In order to recover the exact result, we first use the normal ordered
expansion of $c(\ell)$. With $D(t)\equiv b(t)-b\dag(t)$, where the
time evolution is given by the Heisenberg representation with
$H(\ell=\infty)=\omega b\dag b$, the Green function reads:
\begin{align}
  G(t)&=-i\Theta(t)e^{-\tilde{\lambda}^2} \langle
  c(t)\sum_{n=0}\frac{\tilde{\lambda}^n}{n!}:D^n(t):c\dag
  \sum_{m=0}\frac{\tilde{\lambda}^{m}}{m!}(-1)^{m}:D^{m}(0):\rangle\\
\label{Green_Two_InB}
&=-i\Theta(t)e^{-\tilde{\lambda}^2}
\sum_{n,m=0}\frac{\tilde{\lambda}^n}{n!}\frac{\tilde{\lambda}^{m}}{m!}
(-1)^{m}\langle :D^n(t)::D^{m}(0):\rangle\\\label{Green_Three_InB}
&=-i\Theta(t)e^{-\tilde{\lambda}^2}
\sum_{n,m=0}\frac{\tilde{\lambda}^n}{n!}\frac{\tilde{\lambda}^{m}}{m!}
n!\delta_{n,m}e^{-in\omega t}
\end{align}
To get from Eq. (\ref{Green_Two_InB}) to Eq. (\ref{Green_Three_InB})
we used the following formula (Appendix \ref{AppNormalOrdering}):
\begin{align}
\label{NormalOrdering_InB}
:(b-b\dag)^n::(b-b\dag)^m:=:\exp(\langle(b-b\dag)^2\rangle
\frac{\partial^2}{\partial x_1 \partial
  x_2})x_1^nx_2^{m}|_{x_1=x_2=(b-b\dag)}:\quad,
\end{align}
with $\langle(b-b\dag)^2\rangle=-1$ and
$\langle:(b-b\dag)^n:\rangle=0$ at $T=0$. Summing up the series in Eq.
(\ref{Green_Three_InB}) indeed yields the exact result given in Eq.
(\ref{Green_Function_InB}).

In order to show that also the non-normal ordered expansion of
$c(\ell)$ leads to the correct result, we have to normal order this
expansion. For this we need the following formula (Appendix
\ref{AppNormalOrdering}):
\begin{align}
  (b-b\dag)^n=\sum_{k=0}^{[\frac{n-1}{2}]}\frac{1}{k!}\frac{n!}{2^k(n-2k)!}
  \langle(b-b\dag)^2\rangle^k :(b-b\dag)^{n-2k}:
\end{align}
Considering for the moment only the first $(N+1)$ even powers of
$(b-b\dag)$, we obtain
\begin{align}
  \sum_{n=0}^N\frac{\tilde{\lambda}^{2n}}{2n!}(b-b\dag)^{2n}&=
  \sum_{n=0}^N\sum_{k=0}^n\frac{\tilde{\lambda}^{2k}}{2^k}\frac{G^k}{k!}
  \frac{\tilde{\lambda}^{2(n-k)}}{2(n-k)!}:(b-b\dag)^{2(n-k)}:\\
  &=\sum_{m=0}^N\frac{\tilde{\lambda}^{2m}}{2m!}:(b-b\dag)^{2m}:
  \sum_{k=0}^{N-m}\frac{\tilde{\lambda}^{2k}}{2^k}\frac{G^k}{k!}\quad,
\end{align}
where we introduced $G\equiv\langle(b-b\dag)^2\rangle$,
$\langle...\rangle$ denoting the canonical ensemble average over a
free bosonic system .  The summation of the first $(N+1)$ odd powers
of $(b-b\dag)$ yields
\begin{align}
  \sum_{n=0}^N\frac{\tilde{\lambda}^{2n+1}}{(2n+1)!}(b-b\dag)^{2n+1}=
  \sum_{m=0}^N\frac{\tilde{\lambda}^{2m+1}}{(2m+1)!}:(b-b\dag)^{2m+1}:
  \sum_{k=0}^{N-m}\frac{\tilde{\lambda}^{2k}}{2^k}\frac{G^k}{k!}\quad.
\end{align}
In the limit $N\to\infty$ we obtain
\begin{align}
  \sum_{n=0}\frac{\tilde{\lambda}^n}{n!}(b-b\dag)^n
  =e^{\frac{1}{2}G\tilde{\lambda}^2}
  \sum_{n=0}\frac{\tilde{\lambda}^n}{n!}:(b-b\dag)^n:\quad,
\end{align}
which is an extension of the previous normal ordering of Eq.
(\ref{normal}) to finite temperatures, since $G=-(1+n)$,
$n\equiv(e^{\beta\omega}-1)^{-1}$ being the Bose factor. This shows
that both expansions of $c(\ell)$ are equivalent.

To complete the discussion we will now verify that the
anti-commutation relation $\{c(\ell),c\dag(\ell)\}=1$ holds for all
$\ell$. To show this we will employ the non-normal ordered expansion.
This yields
\begin{align}
  \{c(\ell),c\dag(\ell)\}&=\frac{1}{2}\sum_{n,n'=0}\gamma_n\gamma_{n'}
  ((-1)^n+(-1)^{n'})(b-b\dag)^{n+n'}\\
  &=\sum_{n=0}\sum_{k=0}^{2n}(-1)^k\gamma_{2n-k}\gamma_k(b-b\dag)^{2n}=1\quad.
\end{align}

Summarizing, the series expansion of an operator into bosonic
operators yields consistent results. This is no trivial result since
expanding the bounded operator $c$ into unbounded operators
$(b-b\dag)^n$ might have led to inconsistencies. Further, it has to be
born in mind that the initial operator of the operator flow is
resembled by $c(\ell=0)=c\otimes1_B$, with $1_B$ being the unity
operator of the bosonic Hilbert space. One consequence then is that
the trace of the initial operator is unbounded and thus not defined.

As a second result, we want to mention that both expansions, normal
ordered and non-normal ordered, are equivalent if no approximations
are involved. Nevertheless, the operator expansion into {\em normal
  ordered} operators seems to be a distinguished expansion since it
resembles a non-perturbative approach including the Debye-Waller
factor.

\section{The Constant Term in the Expansion of $\sigma_x$ and $\sigma_z$}
\label{Aconst_f}

In this appendix we want to comment on the constant term $f_i$
appearing in the expansion of the Pauli spin matrices $\sigma_x$ and
$\sigma_z$. This term seems to contradict the theorem of the
invariance of the trace under unitary transformations. But since the
trace of $\sigma_i(\ell=0)=\sigma_i\otimes1$, $1$ being the identity
of the bosonic Hilbert space, does not exist and since we also expand
the Pauli spin matrices in a series of unbounded operators the above
mentioned theorem does not hold anymore. To make sure that the
constant term is indeed physical, one can truncate the Hilbert space
by introducing the ``bosonic'' operator
\begin{align}
  b\to b_N=b\sqrt{(1-b\dag b/N)}\quad,
\end{align}
with $N$ being a positive integer. The truncated Hilbert space is now
only spanned by $N$ vectors
$|\nu\rangle=(b\dag)^{\nu}/\sqrt{\nu!}|0\rangle$ with $\nu=0...N-1$
and $b|0\rangle=0$.  For $N\rightarrow\infty$ we recover the bosonic
Hilbert space.  The above theorem is guaranteed due to the new,
non-canonical commutation relation $[b_N,b_N\dag]=1-(1+2b\dag b)/N$
which obeys the cyclic invariance of the trace:
\begin{align}
  \text{tr}([ b_N,b_N\dag])=\sum_{\nu=0}^{N-1}(1-\frac{1+2\nu}{N})=0
\end{align}
The flow equations now have to be extended to include the flow of the
operator $b\dag b$ that appears in the commutator relation and that
scales as $1/N$. The constant term $f_i$ appears nevertheless and is
governed by the same differential equation as $N\to\infty$. Both terms
together, the constant term $f_i$ and the bosonic bilinear $b\dag b$,
make sure that no trace is generated during the flow.

\section{Upgraded Flow Equations for $\sigma_z$}
\label{HigherOrders}

In this Appendix, we will set up the flow equations for the Pauli matrix $\sigma_z$ including higher orders in the bosonic operators.
Since the basic objects of our expansion are normal ordered operators
we will first give some {(anti-)}commutation relations which are helpful to
evaluate the commutator $[\eta,\sigma_z]$ (see also Appendix
\ref{AppNormalOrdering}):
\begin{align}
  [x,:p^nx^m:]&=-2n:p^{n-1}x^m:\quad,\quad[p,:p^nx^m:]=2m:p^nx^{m-1}:\\
  \{x,:p^nx^m:\}&=2:p^nx^{m+1}:+2m:p^nx^{m-1}:1_n\\
  \{p,:p^nx^m:\}&=2:p^{n+1}x^m:-2n:p^{n-1}x^m:1_n
\end{align}
The commutator of two tensor products of the fermionic and bosonic
Hilbert space can be written as $[oB,o'B']=\{o,o'\}[B,B']/2+
[o,o']\{B,B'\}/2$ which is a useful identity if the anti-commutator
$\{o,o'\}$ vanishes ($o,o'\in\mathcal{H}_e$, $B,B'\in\mathcal{H}_b$).

We will not present the commutator $[\eta,\sigma_z]$ explicitly but
only account for the additional terms that appear with respect to the
previous flow equations. For the first extension $\sigma_z^{new,2}$, this yields:
\begin{align}
\label{gz_Zwei}
\dl g_z&=...+2\eta^x\chi^1\quad,\quad\dl h_z=...+2\eta^z\chi^1\\
\dl \chi^x&=...-2\eta^z\psi^y1_n-4\eta^y\psi^{z,+}1_n\\
\dl \chi^y&=...-4\eta^z\psi^{x,-}1_n+4\eta^x\psi^{z,-}1_n\\
\dl \chi^z&=...+2\eta^x\psi^y1_n+4\eta^y\psi^{x,+}1_n\\
\dl \chi^1&=4\eta^z\psi^{z,+}-2\eta^y\psi^y+4\eta^x\psi^{x,+}\\
\dl \psi^{x,+}&=-2\eta^y\chi^z-2\eta^{0,y}\psi^{z,+}\\
\dl \psi^y&=2\eta^z\chi^x-2\eta^x\chi^z\\
\dl \psi^{z,+}&=2\eta^y\chi^x+2\eta^{0,y}\psi^{x,+}\\
\dl \psi^{x,-}&=2\eta^z\chi^y-2\eta^{0,y}\psi^{z,-}\\\label{psiz_Zwei}
\dl \psi^{z,-}&=-2\eta^x\chi^y+2\eta^{0,y}\psi^{x,-}
\end{align}

Additional contributions relative to the previous flow equations coming
from $\sigma_z^{new,3}$ read:
\begin{align}
\label{chix_Drei}
\dl \chi^x=...+4\eta^x\psi^{1,+}&\;,\; \dl
\chi^y=...-4\eta^y\psi^{1,-}\;,\;
\dl \chi^z=...+4\eta^z\psi^{1,+}\\
\dl \psi^{x,+}&=...-2\eta^z\varphi^{y,+}1_n-6\eta^y\varphi^{z,+}1_n\\
\dl \psi^y&=...-4\eta^z\varphi^{x,-}1_n+4\eta^x\varphi^{z,-}1_n\\
\dl \psi^{z,+}&=...+2\eta^x\varphi^{y,+}1_n+6\eta^y\varphi^{x,+}1_n\\
\dl \psi^{x,-}&=...-6\eta^z\varphi^{y,-}1_n-4\eta^y\varphi^{z,-}1_n\\
\dl \psi^{z,-}&=...+6\eta^x\varphi^{y,-}1_n+4\eta^y\varphi^{x,-}1_n
\end{align}
The flow equations for the new parameters of $\sigma_z^{new,3}$ yield:
\begin{align}
  \dl \psi^{1,+}&=6\eta^z\varphi^{z,+}+2\eta^y\varphi^{y,+}
  +6\eta^x\varphi^{x,+}\\
  \dl \psi^{1,-}&=2\eta^z\varphi^{z,-}+6\eta^y\varphi^{y,-}
  +2\eta^x\varphi^{x,-}\\
  \dl \varphi^{x,+}&=-2\eta^y\psi^{z,+}-2\eta^{0,y}\varphi^{z,+}\\
  \dl \varphi^{y,+}&=2\eta^z\psi^{x,-}-2\eta^x\psi^{z,-}\\
  \dl \varphi^{z,+}&=2\eta^y\psi^{z,+}+2\eta^{0,y}\varphi^{x,+}\\
  \dl \varphi^{x,-}&=2\eta^z\psi^y-2\eta^y\psi^{z,-}-2\eta^{0,y}\varphi^{z,-}\\
  \dl \varphi^{y,-}&=2\eta^z\psi^{x,+}-2\eta^x\psi^{z,+}\\
  \dl
  \varphi^{z,-}&=-2\eta^x\psi^y+2\eta^y\psi^{x,-}+2\eta^{0,y}\varphi^{x,-}
\label{phiz_Drei}
\end{align}

\section{Rabi Model in Perturbation Theory}
\label{RabiPertubation}

In this appendix we will treat the Rabi Hamiltonian in perturbation
theory. We want to start from the exactly solvable Jaynes-Cummings
Hamiltonian which is obtained from the symmetric Rabi Hamiltonian with
no bias by applying the rotating wave approximation \cite{Jay63}. This
approximation neglects coupling or transition terms which are
energetically unlikely.

It is useful to write the Hamiltonian in a basis where $\sigma_x$ is
diagonal.  The Rabi Hamiltonian shall thus be given by
\begin{align}
\label{AppendixRabiHamiltonian}
H=\sum_{i=0,1}\epsilon_ic_i^{\dagger}c_i+\omega_0b^{\dagger}b +\lambda
bc_1^{\dagger}c_0+\lambda b^{\dagger}c_0^{\dagger}c_1
+\lambda'b^{\dagger}c_1^{\dagger}c_0+\lambda'bc_0^{\dagger}c_1\quad.
\end{align}
The operators $c_i^{(\dagger)}$ and $b^{(\dagger)}$ obey the canonical
anti-commutation and commutation relations respectively. We identify
the Rabi Hamiltonian given in Eq. (\ref{RabiHamiltonian}) by setting
$\epsilon_1-\epsilon_0=\Delta_0$ and $\lambda=\lambda'=2\lambda_0$ and
the zero external bias.\\

The Jaynes-Cummings Hamiltonian is obtained by setting $\lambda'=0$ in
Eq. (\ref{AppendixRabiHamiltonian}). We want to treat the so called
off-shell processes, characterized by $\lambda'$, within a systematic
perturbation approach. One way to do so is to consider the Hamiltonian
in the basis $\{|0;2n\rangle|1;2n+1\rangle\}$ and
$\{|0;2n+1\rangle,|1;2n\rangle\}$ where the first quantum number
resembles the fermionic state and the second quantum number the
bosonic state. Since the Hamiltonian is symmetric with respect to
parity the two sets decouple and in the following we will only
consider the first set.

In the above basis, the Hamiltonian is tridiagonal and we define the $n$-dependent matrices
\begin{align}
\begin{split}
  D^{\text{on}}(n)&\equiv
\begin{pmatrix}
  \epsilon_1+2n\omega_0&\sqrt{2n+1}\lambda\\
  \sqrt{2n+1}\lambda&D^{\text{off}}(n+1)
\end{pmatrix}\quad,\\ 
D^{\text{off}}(n+1)&\equiv
\begin{pmatrix}
  \epsilon_0+(2n+1)\omega_0&\sqrt{2n+2}\lambda'\\
  \sqrt{2n+2}\lambda'&D^{\text{on}}(n+1)
\end{pmatrix}\quad.
\end{split}
\end{align}
The determinants can formally be evaluated to yield
\begin{align}
\label{determinants}
\text{det}D^{\text{on}}(n)&=(\epsilon_1+2n\omega_0)\text{det}
D^{\text{off}}(n+1)
-(2n+1)\lambda^2\text{det}D^{\text{on}}(n+1)\\\notag
\text{det}D^{\text{off}}(n+1)&=(\epsilon_0+(2n+1)\omega_0)
\text{det}D^{\text{on}}(n+1)
-(2n+2){\lambda'}^2\text{det}D^{\text{off}}(n+2)\quad.
\end{align}

The matrix $D^{\text{on}}(0)$ resembles the representation of the Rabi
Hamiltonian in the above basis. To determine the eigenvalue $\mu$ of
the matrix up to $O({\lambda'}^2)$ we iterate Eq. (\ref{determinants})
starting with $D^{\text{on}}(0)$:
\begin{align}
\label{eigenvalue}
\begin{split}
  \text{det}(D^{\text{on}}(0)-\mu)\to&\left[(\epsilon_1-\mu)
    (\epsilon_0+\omega_0-\mu)-\lambda^2\right]\\
  \times&\left[(\epsilon_1+2\omega_0-\mu)
    (\epsilon_0+3\omega_0-\mu)-3\lambda^2\right]\text{det}(D^{\text{on}}(2)-
  \mu)\\
  -&(\epsilon_1-\mu)2{\lambda'}^2(\epsilon_0+3\omega_0-\mu)
  \text{det}(D^{\text{on}}(2)-\mu)=0
\end{split}
\end{align}

For the eigenvalues we make the ansatz
$\mu=\mu^{(0)}+{\lambda'}^2\mu^{(1)}$. There is no linear term in
$\lambda'$ since the spectrum of $H$ may not depend on the phase of
the coupling constant.

We now order the eigenvalues as follows: The lowest eigenvalues of
order $O({\lambda'}^0)$, $\mu_{0,\pm}^{(0)}$, are determined by setting the
first factor on the right hand side of Eq. (\ref{eigenvalue}) zero. We
obtain the well-known Jaynes-Cummings result
$\mu_{0,\pm}^{(0)}=\epsilon_0+\omega_0-(\bar{\Delta}\mp R_0)/2$ with the
detuning $\bar{\Delta}\equiv (\epsilon_1-\epsilon_0)-\omega_0$ and the
zeroth Rabi frequency $R_0^2=\bar{\Delta}^2+4\lambda^2$. The first
correction to $\mu_{0,\pm}^{(0)}$ then yields
\begin{align}
\label{KorrekturOne}
\mu_{0,\pm}^{(1)}=\frac{1}{\mp R_0}\frac{\bar{\Delta}\omega_0\pm
  R_0\omega_0-\lambda^2} {2\omega_0^2\mp R_0\omega_0-\lambda^2}\quad.
\end{align}
The result agrees with the perturbative result in the limit
$\lambda=\lambda'\ll\bar{\Delta}$.

Generally, setting the $n$th factor of the first line on the right
hand side of Eq. (\ref{eigenvalue}) zero the $n$th eigenvalues yield
$\mu_{n,\pm}^{(0)}=\epsilon_0+(2n+1)\omega_0-(\bar{\Delta}\mp R_n)/2$ with
$R_n^2=\bar{\Delta}^2+4(2n+1)\lambda^2$. The first correction to
$\mu_{n,\pm}^{(0)}$ is given by
\begin{align}
\begin{split}
\label{KorrekturN}
\mu_{n,\pm}^{(1)}=\frac{1}{\mp R_n}\biggl[(n+1)\frac{\bar{\Delta}\omega_0
  \pm R_n\omega_0-\lambda^2}{2\omega_0^2\mp R_n\omega_0-(n+1)\lambda^2}
+n\frac{\bar{\Delta}\omega_0 \mp
  R_n\omega_0-\lambda^2}{2\omega_0^2\pm
  R_n\omega_0-(n-1)\lambda^2}\biggr] \quad.
\end{split}
\end{align}

The perturbative approach breaks down when degenerated
states are involved. This is indicated by the poles in the energy corrections
$\mu_{n,\pm}^{(1)}$. Setting the denominator of $\mu_{0,\pm}^{(1)}$ zero, 
one obtains for the tunnel-matrix element
$\Delta_0=\omega_0+\sqrt{(2\omega_0^2-\lambda^2)^2-4\omega_0^2\lambda^2}
/\omega_0$.  Inserting the parameters used for Fig.
\ref{EpsilonZero_Rabi}, we obtain $\Delta_0\approx2.87$. This value
approximately agrees with the value of $\Delta_0$ where the second
spike of $h_z$ in Figure \ref{EpsilonZero_Rabi} is seen.
\section{Normal Ordering}
\label{AppNormalOrdering}

In this appendix we want to summarize basic relations concerning
normal ordering. This summary is based on unpublished notes by Wegner
of the year $2000$ in which he presents a general formalism for normal
ordering of classical and quantum fields with respect to a bilinear
Hamiltonian \cite{Weg00}. We will restrain ourselves to the normal ordering of
bosonic quantum fields.

Let $b_k$ be any linear combination of Bose creation and annihilation
operators. The matrix $G$ shall describe the correlations of the
operators $b$ for a Hamiltonian $H$ bilinear in the creation and
annihilation operators: $\langle b_kb_l\rangle=G_{kl}$. The commutator
is given by $[b_k,b_l]=G_{kl}-G_{lk}$. Normal ordering of an operator
$A$ with respect to the Hamiltonian $H$ is now defined by:
\begin{align}
  :1:&=1\\
  :\alpha A(b)+\beta B(b):&=\alpha:A(b):+\beta:B(b):\\
  b_k:A(b):&=:b_kA(b):+\sum_lG_{kl}:\frac{\partial A(b)}{\partial
    b_k}:
\end{align}

The product of $m$ operators $b_{k_i}$ with $i=1..m$ is now obtained
by iterating the third equation. One obtains
\begin{align}
  b_{k_1}b_{k_2}...b_{k_m}=:(b_{k_1}+\sum_{l_1}G_{k_1,l_1}
  \frac{\partial}{\partial
    b_{l_1}})(b_{k_2}+\sum_{l_2}G_{k_2,l_2}\frac{\partial}{\partial
    b_{l_2}})...b_{k_m}:\quad,
\end{align}
which can also be written as
\begin{align}
  b_{k_1}b_{k_2}...b_{k_m}=:\exp(\sum_{kl}G_{kl}\frac{\partial^2}{\partial
    b_k^{left}\partial b_l^{right}})b_{k_1}b_{k_2}...b_{k_m}:\quad.
\end{align}
This is Wick's first theorem \cite{Wic50}. The superscripts $^{left}$ and
$^{right}$ indicate that we always pick a pair of factors $b$ and
perform the derivative $\partial/\partial b_k$ on the left factor and
the derivative $\partial/\partial b_l$ on the right factor, so that
the factor $G_{kl}$ depends on the sequence of the operators.

Similarly one obtains
\begin{align}
  :b_{k_1}b_{k_2}...b_{k_m}:=\exp(-\sum_{kl}G_{kl}\frac{\partial^2}{\partial
    b_k^{left}\partial b_l^{right}})b_{k_1}b_{k_2}...b_{k_m}\quad.
\end{align}

The formula for the product of two normal ordered operators is given
by
\begin{align}
\label{WickTheorm_Boson}
:A(b)::B(b):=:\exp(\sum_{kl}G_{kl}\frac{\partial^2}{\partial
  b_k\partial a_l})A(b)B(a):|_{a=b}\quad.
\end{align}
This is Wicks's second theorem.

One can now show that under normal ordering the commutative law holds:
${:ABCD:=:ACBD:}$. This is rule C of Wick.

%

%


\begin{thebibliography}{99}

%
\bibitem{Gla94} S.D. G{\l}azek and K.G. Wilson, Phys. Rev. D \textbf{48}, 5863 
  (1993) and Phys. Rev. D \textbf{49}, 4214 (1994). 
\bibitem{Weg94}
  F. Wegner, Ann.~Phys. (Leipzig) \textbf{3}, 77 (1994).  	
\bibitem{Keh96a} S.K. Kehrein and A. Mielke,  Phys. Lett. A
 \textbf{219}, 313 (1996).  
\bibitem{Keh96b} S.K. Kehrein, A. Mielke, and P. Neu, Z. Phys. B
  \textbf{99}, 269 (1996).  
\bibitem{Keh97} S.K. Kehrein and A. Mielke, Ann.~Phys. (Leipzig) \textbf{6},
  90 (1997).      
\bibitem{Len96}
P. Lenz and F. Wegner, 
 Nucl. Phys. B \textbf{482} [FS], 693 (1996). 
\bibitem{Rag99} M. Ragwitz and F. Wegner,
  Eur. Phys. J. B \textbf{8}, 9 (1999). 
\bibitem{Gro02}
I. Grote, E. Koerding, and F. Wegner, 
J. Low Temp. Phys. \textbf{126}, 1385 (2002).  
\bibitem{Weg01} F.~Wegner, Physics Reports {\bf348}, 77 (2001).
\bibitem{Sas90}
M. Sassetti and U. Weiss, Phys. Rev. Lett. \textbf{65}, 2262 (1990).
\bibitem{Ric97} J. Richter, Diploma thesis, University of Heidelberg, (1997).
\bibitem{Leg87}
  A.J. Leggett, S. Chakravarty, A.T. Dorsey, M.P.A. Fisher, A. Grag, and W. Zwerger, Rev. Mod. Phys. \textbf{59}, 1 (1987); erratum. Rev. Mod. Phys.
  \textbf{67}, 725 (1995). 
\bibitem{GH84} R. Graham and M. H\"ohnerbach, 
  Z. Phys. B {\bf 57}, 233 (1984).
\bibitem{Cibils91} M.B. Cibils, Y. Cuche, V. Marvulle, W.F. Wreszinski, 
  J.-P. Amiet, and H. Beck, 
  J. Phys. A {\bf 24}, 1661 (1991).
\bibitem{Cibils92} M.B. Cibils, Y. Cuche, P. Leboeuf,  and W.F. Wreszinski,
  Phys. Rev. A {\bf 46}, 4560 (1992).
\bibitem{Sta00}
T. Stauber, R. Zimmermann, and H. Castella, Phys. Rev. B \textbf{62}, 7336 (2000).
\bibitem{Mielke98} A. Mielke, Eur. Phys. J. B {\bf 5}, 605 (1998).   
\bibitem{close} In Appendix \ref{IndependentBoson} we show that an
  infinite series of newly generated operators can be summed up for an
  exactly solvable model to yield a closed expression.
\bibitem{Jay63} E.T. Jaynes and F.W. Cummings,
  Proc. IEEE \textbf{51}, 89 (1963). 
\bibitem{StaDiss} T. Stauber, PhD thesis, University of Heidelberg, (2002).
\bibitem{Sta02} T. Stauber and A. Mielke, cond-mat 0207414, to appear in Phys. Lett. A.
\bibitem{error} We neglect errors resulting from the truncation
  of the Hilbert space. 
\bibitem{Reso} In Appendix \ref{RabiPertubation}
  the Rabi model is treated in perturbation theory based on the
  exactly solvable Jaynes-Cummings Model. In this context resonances
  are characterized by the vanishing of the energy denominator of the
  perturbative expansion.
\bibitem{Keh00} S.K. Kehrein, Nucl. Phys. B \textbf{592} [FS], 512 (2001).
\bibitem{Mah90} G.D.L. Mahan, {\it Many-Particle
    Physics}. (Plenum, New York, 1990).
\bibitem{Weg00} F. Wegner, unpublished.
\bibitem{Wic50}
G.C. Wick, Phys. Rev. \textbf{80}, 268 (1950). 
\end{thebibliography}
\end{document}